\documentclass[10pt,a4paper,fleqn]{article}
\usepackage{amsmath}
\usepackage{amssymb}
\usepackage{amscd}
\usepackage[dvips]{graphicx}

\newtheorem{theorem}{Theorem}

\newtheorem{corollary}[theorem]{Corollary}
\newtheorem{definition}[theorem]{Definition}
\newtheorem{example}[theorem]{Example}
\newtheorem{lemma}[theorem]{Lemma}
\newtheorem{proposition}[theorem]{Proposition}
\newtheorem{remark}[theorem]{Remark}
\newtheorem{conjecture}[theorem]{Conjecture}

\newenvironment{proof}{{\bf Proof. }}{\hfill$\rule{1ex}{1ex}$\par\medskip}

\newcommand{\bt}{\begin{theorem}}
\newcommand{\et}{\end{theorem}}
\newcommand{\bd}{\begin{definition}}
\newcommand{\ed}{\end{definition}}
\newcommand{\bs}{\begin{proposition}}
\newcommand{\es}{\end{proposition}}
\newcommand{\bp}{\begin{proof}}
\newcommand{\ep}{\end{proof}}
\newcommand{\be}{\begin{equation}}
\newcommand{\ee}{\end{equation}}

\newcommand{\br}{\begin{remark}}
\newcommand{\er}{\end{remark}}
\newcommand{\bex}{\begin{example}}
\newcommand{\eex}{\end{example}}
\newcommand{\bc}{\begin{corollary}}
\newcommand{\ec}{\end{corollary}}
\newcommand{\bl}{\begin{lemma}}
\newcommand{\el}{\end{lemma}}
\newcommand{\bj}{\begin{conjecture}}
\newcommand{\ej}{\end{conjecture}}

\def\tc{{\tt c}}
\def\td{{\tt r}}

\def\tM{{\tt M}}

\def\tR{{\tt R}}

\def\tS{{\tt S}}
\def\ts{{\tt s}}
\def\tT{{\tt T}}
\def\ttt {{\tt t}}

\def\tU{{\tt U}}

\def\tx{{\tt x}}
\def\ty{{\tt y}}
\def\t0{{\tt 0}}

\def\vphi{\varphi}

\def\orho{{\overline\rho}}
\def\oUps{\overline{\Upsilon}}
\def\obUps{\overline{\mbox{\boldmath$\Upsilon$}}}

\def\om{\omega}
\def\oom{\overline\omega}

\def\bom{{\mbox{\boldmath$\omega$}}}
\def\obom{\overline\bom}
\def\obOm{\overline\bOm}
\def\bOm{\underline\bOm}

\def\0bom{{\bom}^0}
\def\0obom{{\obom}^0}

\def\0nbom{{\bom}_{n,0}}
\def\n*bom{{\bom}^*_{(n)}}
\def\wt{\widetilde}

\def\oom{\overline\om}

\def\Om{\Omega}
\def\oOm{\overline\Om}
\def\obOm{\underline\Om}
\def\bOm{\mbox{\boldmath${\Om}$}}
\def\obOm{\overline\bOm}

\def\gam{\gamma}
\def\Lam{\Lambda}
\def\Lamc{\Lam^{\complement}}
\def\LamR{\Lam^{(\tR)}}

\def\0Lamc{\Lam_0^{\rm c}}

\def\ups{\upsilon}

\def\bbC{\mathbb C}

\def\fB{\mathfrak B}

\def\fF{\mathfrak F}

\def\fK{\mathfrak K}
\def\fM{\mathfrak M}
\def\fS{\mathfrak S}

\def\fW{\mathfrak W}

\def\dist{\rm{dist}}

\def\cC{\mathcal C}
\def\cD{\mathcal D}

\def\cG{\mathcal G}
\def\cH{\mathcal H}
\def\cJ{\mathcal J}
\def\cK{\mathcal K}

\def\cW{\mathcal W}

\def\ocW{\overline\cW}
\def\ucW{\underline\cW}

\def\bbC{\mathbb C}
\def\bbP{\mathbb P}
\def\obbP{\overline\bbP}
\def\ubbP{\underline\bbP}

\def\bbR{\mathbb R}

\def\bbZ{\mathbb Z}

\def\bfI{\textbf I}

\def\bx{\mathbf x}
\def\by{\mathbf y}
\def\obx{\overline\bx}

\def\oby{\overline\by}

\def\unn{{\underline{n_{}}}}
\def\ux{{\underline{x_{}}}}

\def\by{\mathbf y}
\def\uy{\underline y}

\def\ox{\overline x}
\def\obx{\overline\bx}

\def\dist{\textrm{dist}}
\def\diy{\displaystyle}
\def\ov{\overline}

\def\uy{\underline y}

\def\uz{\underline{z_{}}}

\def\om{\omega}
\def\oom{\ov\om}

\def\bI{\mathbf I}
\def\bII{\mathbf{II}}

\def\ocW{\overline\cW}

\def\rc{{\rm c}}

\def\rd{{\rm d}}

\def\ru{{\rm u}}
\def\rv{{\rm v}}
\def\rx{{\rm x}}
\def\ry{{\rm y}}
\def\tr{{\rm{tr}}}

\def\obZ{{\overline{\mathbf Z}}}
\def\orZ{{\overline{\rm Z}}}
\def\rmm{\rm m}

\begin{document}
\title{\bf FK-DLR properties\\
of a quantum multi-type bose-gas\\
with a repulsive interaction}
\author{\bf Y. Suhov~$^{1}$, I. Stuhl$^{2}$}
\vspace{1mm}\maketitle {\footnotesize

\noindent $^1$ Statistical Laboratory, DPMMS, University of Cambridge, UK;\\
Department of Statistics/IME,  University of S\~ao Paulo, Brazil;\\
IITP, RAS, Moscow, Russia\\
E-mail: yms@statslab.cam.ac.uk

\noindent $^2$ University of Debrecen, Hungary\\
IME, University of S\~ao Paulo, Brazil;\\
E-mail: izabella@ime.usp.br}

\begin{abstract}
The paper extends earlier results from \cite{SK}, \cite{SKS} about infinite-volume quantum
bosonic states (FK-DLR states) to the case of multi-type particles with non-negative interactions.
(An example is a quantum Widom--Rowlinson model.)
Following the strategy from  \cite{SK}, \cite{SKS},
we establish that, for the values of fugacity $z\in (0,1)$ and inverse temperature $\beta >0$,
finite-volume Gibbs states form a compact family in the thermodynamic limit. Next,  in
dimension two we show that any limit-point state (an FK-DLR state in the terminology
adopted in \cite{SK}, \cite{SKS}) is translation-invariant.
\\ \\
\textbf{2000 MSC:} 60F05, 60J60, 60J80.\\
\vskip.1truecm

\textbf{Keywords:} bosonic quantum multi-type system in $\bbR^d$,
Hamiltonian, Laplacian, two-body interaction,
finite-range potential, Fock space,
Gibbs operator, FK-representation,
density matrix, Gibbs state, reduced density matrix, thermodynamic limit,
FK-DLR equations, FK-DLR states and functionals
\end{abstract}

\section{Limit-point Gibbs states and reduced density matrices}

The present paper is a continuation of earlier works \cite{SK}, \cite{SKS}. As in
\cite{SK}, \cite{SKS}, we attempt at establishing a working definition of an
infinite-volume quantum bosonic Gibbs state and justify it by checking
natural properties such as shift-invariance in dimension two. In addition, the paper
lays a foundation for future research into phase transitions in quantum Widom--Rowlinson
(WR) models
with several types of particles (following a recent progress in classical WR models; see \cite{MSS1, MSS2}). Cf. the earlier work \cite{CP}.
The class of states
under consideration is formed by the so-called FK-DLR states (a more general concept is
an FK-DLR functional): these states satisfy a quantum analog of the DLR equation (after
Dobrushin--Lanford--Ruelle). Throughout the paper,
we refer to \cite{SK} and \cite{SKS} by adding the Roman numerals ${\bI}$ and ${\bII}$,
respectively: Theorem 1.2.${\bI}$, formula (4.1.${\bII}$), etc.

The difference between the present work and \cite{SK}, \cite{SKS} is in assumptions
upon the interaction potential, which implies different conditions on the thermodynamic
variables $z$ (the fugacity) and $\beta$ (the inverse temperature). Besides, in this  paper we
consider systems with several particle types $i\in\{1,\ldots ,q\}$. We suppose that the
(two-body) interaction potentials $V_{ij}$ between types $i,j$ are non-negative (i.e., generate
a repulsion of particles); they may also include hard cores. The non-negativity assumption allows
us to work in the open domain $z_1,\ldots ,z_q\in (0,1)$, $\beta >0$: in the border of this domain
(where $z_j=1$ for some $j$) one may expect a Bose--Einstein condensation (which occurs
for $V_{ij}\equiv 0$). However, even for $z_j$ less than (but close to) $1$ one can't exclude
(at least at a rigorous level) a non-uniqueness of an infinite-volume Gibbs state as it has been
defined in \cite{SK}, \cite{SKS} and in the current paper. In the quantum bosonic WR model with
$z_1=\ldots =z_q=z$ we expect a first-order phase transition for $z\in (1-\eta ,1)$ (cf.
\cite{CP}).
\medskip

{\bf Remark} Historically, the assumption $V\geq 0$ was used, elegantly and to a great effect, by Ginibre in \cite{Gi1} and became
popular in quantum Statistical Mechanics (cf. the references \cite{BR1}, \cite{Su2}, to name a few).
Admittedly, this assumption was termed in \cite{Gi1} ``a severe physical limitation'', and it was
declared that the ``next task is to get rid of it''. To a certain extent, it was achieved in \cite{Gi2}.
Indeed, it can be noted that in  \cite{Gi2} and \cite{Gi3} (where a number of different
conditions upon the potential were introduced
and intermittently used) the assumption of non-negativity was not present. However
(and perhaps, consequently), the
conditions upon $z$ and $\beta$ guaranteeing the key result of \cite{Gi2}, \cite{Gi3}  (convergence to a unique
infinite-volume limit and cluster expansion of the quantum Gibbs state) became notably less transparent than
in \cite{Gi1}.
\medskip

The present paper follows the approach adopted in \cite{SK}, \cite{SKS}; this allows us to use
pre-requisites and technical tools from the above references. However, we attempted at making
this paper, to a degree, self-sufficient, as far as the statements of the main theorems are concerned.
In Sections 1.1, 1.2 and 1.3 we introduce the models,
state the main results and discuss the principal tool of the work: the Feynman--Kac (FK)
representation. In Sections 2.1, 2.2 and 2.3 we prove the existence
of an infinite-volume FK-DLR state). Finally, in Section 3 we focus on the 2D case and check that
any FK-DLR functional is shift-invariant.

\subsection{The local Hamiltonian}

A  model of a
quantum Bose-gas in $\bbR^d$ with $q$ types of particles is determined by a family of local
Hamiltonians. Given a vector $\unn =(n(1),\ldots ,n(q))$ with non-negative integer entries $n(j)$.
Consider a system with $n(j)$ particles of type $j\in\{1,\ldots ,q\}$ in a finite `box' (a
$d$-dimensional cube)
$$\Lambda (=\Lam_L) =[-L,+L]^{\times d}.$$
The Hamiltonian, $H_{\unn,\Lam}$, is a self-adjoint operator acting on functions
$\phi_{\unn}\in\cH_\unn (\Lam ):=\operatornamewithlimits{\otimes}\limits_{1\leq j\leq q}{\rm L}_2^{{\rm{sym}}}
(\Lam^{n(j)})$:
$$\begin{array}{l}
\diy\left(\,H_{\unn,\Lam}\phi_{\unn}\right)\left(\ux^{\unn}\,\right)
=-\frac{1}{2}\;\sum\limits_{1\leq j\leq q}\;\sum\limits_{1\leq l\leq n(j)}
\left(\Delta_{j,l}\phi_\unn\right)\left(\ux^\unn\right)\\
\qquad\diy +\sum\limits_{1\leq j\leq j^\prime\leq q}\;\sum\limits_{1\leq l\leq n(j)}\;
\sum\limits_{1\leq l^\prime\leq n(j^\prime )}
V_{j,j^\prime}\left(\left|x_{j,l}-x_{j^\prime ,l^\prime}\right|\right)
\phi_\unn\left(\ux^\unn\right),\\
\qquad\qquad\ux^\unn=(\ux(1),\ldots ,\ux(q)),\;\ux (j)=(x_{j,1},\ldots ,x_{j,n(j)})\in\Lam^{n(j)}.
\end{array}\eqno (1.1.1)$$
Here function $\phi_\unn$ is symmetric under permutations of variables $x_{j,l}$ within
each group $\ux (j)$. The symbol $\left|x_{j,l}-x_{j^\prime ,l^\prime}\right|$ stands for the
Euclidean distance between points $x_{j,l},x_{j^\prime ,l^\prime}\in\bbR^d$. (Sometimes we use an
alternative notation $x(j,l),x(j^\prime,l^\prime )$.)

Operator $\Delta_{j,l}$ in (1.1.1) acts as a Laplacian in the variable $x_{i,l}$.
Further, $V_{j,j^\prime}:\,r\in [0,+\infty )\mapsto V_{j,j^\prime}(r)\in [0,+\infty ]$ is assumed to
be a $C^2$-function at each point where $V_{j,j^\prime}<+\infty$ and with a compact support.
Function $V_{j,j^\prime}$ describes a two-body interaction potential between type $j$ and
type $j^\prime$ particles, depending upon the distance between the particles involved. The
value
$$\tR=\inf\,\Big[r>0:\;V_{j,j^\prime}({\wt r})\equiv 0\;\hbox{ for }\;{\wt r}
\geq r,\;1\leq j\leq j^\prime\leq q\,\Big]\eqno (1.1.2)$$
is called the interaction radius (or the interaction range). We also assume that $V_{j,j^\prime}(r)$
may take the value $+\infty$ (e.g., when $0\leq r\leq D(j,j^\prime )$, with $D(j,j^\prime )=
D(j^\prime ,j)\in [0,+\infty )$ representing the diameter of a hard-core repulsion between
particles of types $j$ and $j^\prime$).$^{1)}$\footnote{$^{1)}$In the case $V_{j,j^\prime}=+\infty$ the operator
$H_{\unn,\Lam}$ acts on functions $\phi_{\unn}$ vanishing whenever $V_{J,j^\prime}(\left|x_{j,l}
-x_{j^\prime ,l^\prime}\right|)=+\infty$
(Dirichlet`s boundary conditions).} Next, we suppose, for definiteness, that
$$\begin{array}{l}{\ov V}^{\,(0)}=\max\,\Big[V_{j,j^\prime}(r):\;0\leq r\leq \tR,\\ \qquad\qquad\qquad{}
 V_{j,j^\prime}(r)<+\infty ,\;1\leq j\leq j^\prime\leq q\Big]<+\infty,\end{array}\eqno (1.1.3)$$
and
$$\begin{array}{l}{\ov V}^{\,(1)}=\max\,\Big[\left| V^\prime_{j,j^\prime}(r)\right|:
0\leq r\leq\tR,\\
\qquad\qquad\qquad{}
 V_{j,j^\prime}(r)<+\infty ,\;1\leq j\leq j^\prime\leq q\Big]<+\infty,\\
{\ov V}^{\,(2)}=\max\,\Big[\left| V^{\prime\prime}_{j,j^\prime}(r)\right|: 0\leq r\leq\tR,\\
\qquad\qquad\qquad{}
 V_{j,j^\prime}(r)<+\infty ,\;1\leq j\leq j^\prime\leq q\Big]<+\infty .
\end{array}\eqno (1.1.4)$$

{\bf Remark} As was said above, the assumption that $V_{j,j^\prime}\geq 0$ means that the
interaction potential generates repulsion between particles.  Such a condition was repeatedly
used in the works on quantum systems of Statistical Mechanics; see, e.g., \cite{Gi1}, \cite{Su2},
\cite{BR1}. It covers the case of a free gas (where $V_{j,j^\prime}\equiv 0$). Removing the
non-negativity assumption (without introducing a hard-core of a positive diameter) represents
some challenges and remains an open question. On the other hand, the finite range assumption
is used in this paper for simplifying some technicalities and can be relaxed to a controlled
decay of $V_{j,j^\prime}(r)$ for large $r$; this will be the subject of a forthcoming research.
\medskip

Operator $H_{\unn,\Lam}$ is determined by a boundary condition on $\partial\Lam^\unn$.
Here  $\Lam^\unn =\operatornamewithlimits{\times}\limits_{1\leq j\leq q}\Lam^{n(j)}$ and
$$\begin{array}{l}\partial\Lam^\unn
=\bigg\{\ux^\unn=\big(\ux(1),\ldots ,\ux(q)\big)\in\Lam^\unn:\\
\qquad\qquad{} \max\Big[\big|x_{jl}\big|_{\rmm}:\;1\leq j\leq q,\;1\leq l\leq n(j)\Big]=L\bigg\}
.\end{array}\eqno (1.1.5)$$
Here $|\;\;|_{\rm m}$ stands for the maximum norm in ${\bbR}^d$.
More precisely, we initially consider $H_{\unn,\Lam}$
as a symmetric operator given by the RHS of Eqn (1.1.1)
on the set of C$^2$-functions $\phi =\phi_\unn$ vanishing in a neighborhood of $\partial\Lam^\unn$,
i.e., have the support within the interior of $\Lam^\unn$.
It is a self-adjoint extension of this symmetric operator (denoted by the same
symbol $H_{\unn,\Lam}$) which requires boundary conditions.
Throughout the paper we consider Dirichlet's boundary condition:
$$\phi_\unn (\ux^\unn)=0,\;\;\ux^\unn\in\partial\Lam^\unn .\eqno (1.1.6)$$
Nevertheless, the methods of this paper allow us to consider a broad class of
conditions, viz., elastic boundary conditions (where
a linear combination of the value of $\phi$  at the boundary
and the value of its normal derivative vanishes);
periodic boundary conditions can also included.  Considering various boundary
 conditions endeavors towards including possible phase transitions; this
 question can be left for forthcoming works.

Under the above assumptions, $H_{\unn,\Lam}$ is a self-adjoint operator, bounded
from below and with a pure point spectrum. Moreover, $\forall$ $\beta\in (0,+\infty )$,
the operator $G_{\beta ,\unn,\Lam}=\exp\,\left[-\beta H_{\unn,\Lam}\right]$ (the
Gibbs operator in  $\cH_\unn (\Lam )$ at the inverse temperature $\beta$) is a
positive-definite operator in $\cH_\unn (\Lam)$,  of the trace class. The trace
$$\Xi (\beta,\unn,\Lam ):={\tr}_{\cH_\unn (\Lam)}\;G_{\beta ,\unn,\Lam}
\in (0,+\infty )\eqno (1.1.7)$$
represents the $\unn$-particle partition function in $\Lam$.

As in \cite{SK}, \cite{SKS}, we work with the grand
canonical Gibbs ensemble. Namely, we consider, $\forall$ vector $\uz=(z_1,\ldots ,z_q)\in (0,1 )^q$,
the direct sum
$$G_{\uz,\beta ,\Lam}=\operatornamewithlimits{\oplus}\limits_{\unn\geq 0}
G_{\beta ,\unn,\Lam}\;\prod\limits_{1\leq j\leq q}z_j^{n(j)}.\eqno (1.1.8)$$
This determines a positive-definite trace-class operator $G_{\uz,\beta ,\Lam}$
in the bosonic Fock space
$$\cH (\Lam )=\operatornamewithlimits{\oplus}\limits_{\unn\geq 0}\;
\cH_\unn (\Lam ).\eqno (1.1.9)$$
The quantity
$$\Xi_{\uz,\beta }(\Lam ):=\sum_{\unn\geq 0}\Xi (\beta,\unn,\Lam )\;\prod\limits_{1\leq j\leq q}z_j^{n(j)}
={\tr}_{\cH (\Lam )}G_{\uz ,\beta ,\Lam}\in (0,+\infty ) \eqno (1.1.10)$$
yields the grand canonical partition function in $\Lam$ at fugacity
$\uz$ and the inverse temperature $\beta$. Further, the operator
$$R_{\uz,\beta ,\Lam}=\frac{1}{\Xi_{\uz,\beta}(\Lam )}
G_{\uz,\beta,\Lam}\eqno (1.1.11)$$
is called the (grand-canonical) density matrix (DM) in $\Lam$;
this is a positive-definite operator in $\cH (\Lam )$ of trace $1$.
Operator $R_{\uz,\beta ,\Lam}$ determines the Gibbs state (GS),
i.e., a linear positive normalized functional $\varphi_{\uz,\beta ,\Lam}$
on the C$^*$-algebra $\fB(\Lam )$ of bounded operators in $\cH (\Lam )$:
$$\varphi_{\uz,\beta ,\Lam}(A)={\tr}_{\cH (\Lam )}\big(AR_{\uz,\beta ,\Lam}\big),
\;\;A\in\fB(\Lam ).\eqno (1.1.12)$$
\medskip
{\bf Remark} The assumption that $0<z_j<1$ means that we avoid
a (possible) `critical` regime. Viz., it is the values $z_j\nearrow 1$ that generates
a Bose-condensation in the free gas ($V_{j,j^\prime}\equiv 0$) in dimension $d\geq 3$. Cf. \cite{BR2}
and the references therein.
\medskip

As in \cite{SK}, \cite{SKS}, the object of interest in this paper is the reduced DM
(briefly, RDM), in cube $\Lam_0\subset\Lam$ centered at a point $c_0=(\tc_0^1,\ldots ,\tc_0^d)$:
$$\Lam_0=[-L_0+\tc_0^{1},\tc_0^{1}+L_0]\times\cdots\times [-L_0+\tc_0^{d},\tc_0^{d}+L_0].
\eqno (1.1.13)$$
As in the aforementioned papers, the term RDM is used here for the operator
$R^{\Lam_0}_{\uz,\beta ,\Lam}$ defined via partial trace$^{2)}$\footnote{$^{2)}$
Our definition the RDM is different from (although close to) that used in \cite{Gi1}, \cite{Gi2}.}
$$R^{\Lam_0}_{\uz,\beta ,\Lam}=\tr_{\cH (\Lam\setminus\Lam_0)}R_{\uz,\beta,\Lam};
\eqno (1.1.14)$$
it is based on the tensor-product representation $\cH (\Lam )=\cH(\Lam_0)\otimes
\cH  (\Lam\setminus\Lam_0)$. Operator $R^{\Lam_0}_{\uz,\beta ,\Lam}$ acts in $\cH(\Lam_0)$,
is positive-definite
and has trace $1$. Furthermore, the partial trace operation generates an important
compatibility property for RDMs. Suppose cubes $\Lam_1\subset\Lam_0\subset\Lam$, then
$$R^{\Lam_1}_{\uz,\beta ,\Lam} =\tr_{\cH (\Lam_0\setminus\Lam_1)}R^{\Lam_0}_{\uz,\beta ,\Lam} .
\eqno (1.1.15)$$
Throughout this paper, we use the upper indices $\Lam_0$ and $\Lam_1$ to indicate
the corresponding `volumes'  have been not affected by the partial trace.

To shorten the notation, the indices/arguments
$\uz$ and $\beta$ will be omitted whenever it does not produce a confusion.
A straightforward modification of the above concepts emerges by including
an external potential field induced by an external classical multi-type configuration (CC)
$\bx(\LamR)$. Such a configuration is represented by a collection $(\bx (\LamR,1),\ldots ,\bx (\LamR,q))$ of finite subset
in an `external` annulus encircling cube $\Lam$:
$$\LamR(=\LamR_L )=\{x\in\bbR^d\setminus\Lam :
\;\dist (x,\Lam)\leq\tR\}.\eqno (1.1.16)$$
Viz., the Hamiltonian $H_{\unn,\Lam |\bx(\LamR)}$ is given by
$$\begin{array}{l}\diy\left(H_{\unn,\Lam |\bx(\LamR)}\phi_\unn\right)\left(\ux^\unn\right)=\left(H_{\unn,\Lam}\phi_\unn
\right)(\ux^\unn)\\
\diy\qquad +\sum\limits_{1\leq j\leq q}\sum\limits_{1\leq l\leq n(j)}\sum\limits_{1\leq j^\prime\leq q}\,\sum\limits_{\ox\in
\bx (\LamR,j^\prime )}V_{j,j^\prime}\left(\left|x_{j,l}-\ox\right|\right)
\phi_\unn\left( \ux^\unn\right)\end{array}\eqno (1.1.17)$$
and has all properties that have been listed above for  $H_{\unn,\Lam}$. This enables
us to introduce the Gibbs operators
$G_{\unn,\Lam |\bx(\LamR)}$ and $G_{\Lam |\bx(\LamR)}$,
the partition functions $\Xi_{\unn}(\Lam |\bx(\LamR))$ and $\Xi (\Lam |\bx(\LamR))$,
the DM $R_{\Lam |\bx(\LamR)}$, the GS $\varphi_{\Lam |\bx(\LamR)}$ and
the RDMs $R^{\Lam_0}_{\Lam |\bx(\LamR)}$ where $\Lam_0\subset\Lam$. Viz.,
$$\begin{array}{c}G_{\unn,\Lam |\bx(\LamR)}=\exp\,\left[-\beta H_{\unn,\Lam |\bx(\LamR)}\right],\;
G_{\Lam  |\bx(\LamR )}=\operatornamewithlimits{\oplus}\limits_{\unn\geq 0}G_{\unn,\Lam  |\bx(\LamR)}\prod\limits_{1\leq j\leq q}z_j^{n(j)},\\ \;\\
 \Xi (\unn,\Lam |\bx(\LamR)):={\tr}_{\cH_\unn (\Lam)}G_{\unn,\Lam |\bx(\LamR},\\ \;\\
\Xi (\Lam |\bx(\LamR)):=\sum\limits_{\unn\geq 0}\Xi (\unn,\Lam |\bx(\LamR))\prod\limits_{1\leq j\leq q}z_j^{n(j)}
={\tr}_{\cH (\Lam )}G_{\Lam |\bx(\LamR)},\\
\diy R_{\Lam |\bx(\LamR)}=\frac{G_{\Lam |\bx(\LamR)}}{\Xi (\Lam |\bx(\LamR))}
\,,\;R^{\Lam_0}_{\Lam |\bx(\LamR)}=\tr_{\cH (\Lam\setminus\Lam_0)}
R_{\Lam |\bx(\LamR)},\\ \;\\
\diy \varphi_{\Lam |\bx(\LamR)}(A)={\tr}_{\cH (\Lam )}\big(AR_{\Lam |\bx(\LamR)}\big),
\;\;A\in\fB(\Lam ).\end{array}\eqno (1.1.18)$$
The previous definitions (1.1.1)--(1.1.15) correspond to the case of an empty exterior CC
$\bx(\LamR )=\emptyset$.

As in \cite{SK}, \cite{SKS}, the Fock spaces $\cH(\Lam )$ and
$\cH (\Lam_0)$ (see (1.1.9)) will be
represented as ${\rm L}_2(\cC (\Lam ))$ and ${\rm L}_2(\cC (\Lam_0))$, respectively.
Here and below, $\cC (\Lam )$ denotes the space formed by collections $\obx =\{\bx_1,\ldots ,\bx_q\}$
of finite (unordered) subsets
$\bx (j)\subset\Lam$ (including the empty set) with the Lebesgue--Poisson measure
$$\begin{array}{l}\diy{\rd}\obx =\prod\limits_{1\leq j\leq q}\frac{1}{(\sharp\;\bx (j))!}\prod_{x\in\bx (j)}{\rd}x,\;\;
\Big(\hbox{with $\diy\int\limits_{\cC (\Lam)}{\rd}\obx
=\exp\,[q\ell (\Lam)]$}\\
\qquad\qquad\qquad\qquad\hbox{where $\ell$ is the Lebesgue measure on $\bbR^d$}\Big).
\end{array}\eqno (1.1.19)$$
Here and later on, the symbol $\sharp$ is used for the cardinality of a given set. In accordance
with this notation, we write that $\bx(\LamR)\in\cC (\LamR)$.
Points $\obx$, $\obx'$, $\bx(\LamR )$ (and $\oby$ later on) are called, as before, classical
multi-type configurations (CCs). We also introduce the subset $\cC(\Lam ,\unn)$ formed by CCs $\obx\in\cC(\Lam )$
with $\sharp\,\bx (j) =n(j)$.
Next, the external CCs $\bx(\LamR)$ have to be controlled, up to a degree, as $\Lam\to\bbR^d$;
see below. The methods
developed in this paper allow us to introduce several methods of such control.  Throughout the
paper we will
refer to the following condition upon a family $\{\bx(\LamR )\}$: for given $(z_1,...,z_q)\in (0,1)^q$ and $\beta >0$,
$\forall$ constant $\rc\in (0,+\infty )$, the quantity
$$B (\rc):=\sup\left[\sum\limits_{1\leq i\leq q}\sharp\,\bx (\LamR_L,i)\sum\limits_{k\geq 1}z_i^kk\,\exp
\,\left(-\frac{L^2-\rc L}{2\beta k}\right):\;L\geq 1\right]<\infty .\eqno (1.1.20) $$
This assumption will be used without stressing it every time again. However, we do not consider
it as a final one; in our opinion, it can be weakened.

\subsection{The thermodynamic limit and the shift-invariance\\ property in two dimensions}

The thermodynamic limit is the key concept of rigorous Statistical Mechanics; in the context of this work
it is $\lim\limits_{\Lam\nearrow\bbR^d}$, the family of standard cubes ordered by inclusion.
In the literature, the quantities and objects identified as limiting
points in the course of this limit are often referred to as infinite-volume ones (e.g., an infinite-volume
RDM or GS). Traditionally, the existence and uniqueness of such a limiting object is treated as absence
of a phase transition. On the other hand, a multitude of such objects (viz., depending on the
boundary conditions for the Hamiltonian or the choice of external CCs) is considered as a
manifestation of a phase transition.

However, since late 1960s there is known an elegant alternative where infinite-volume objects are identified
in terms that do not explicitly invoke the thermodynamic limit. For classical
systems, this is the DLR equations and for so-called quantum spin systems -- the KMS
boundary condition.
The latter is not applicable to the class of quantum systems
under consideration, since the Hamiltonians $H_{\unn,\Lam}$ and
$H_{\unn,\Lam |\bx(\LamR )}$ are not bounded.

In this paper we propose a construction generalising the classical DLR equation (see Section 2.3).
A justification of this construction is given in Section 3 where we
establish the shift-invariance property for the emerging objects (the RDMs and GSs) in dimension two
(i.e., for $d=2$).

\noindent
The first result claimed in this work is

\medskip
{\bf Theorem 1.1.} {\sl Given $\beta\in (0,+\infty )$ and $\uz\in (0,1)^q$, $\forall$ cube $\Lam_0$
(see Eqn {\rm{(1.1.13)}}), the
family of RDMs $\left\{R^{\Lam_0}_{\Lam |\bx(\LamR)},\Lam\nearrow\bbR^d\right\}$
is compact in the trace-norm operator topology in $\cH (\Lam_0)$, for any choices of
CCs $\bx(\LamR)\in\cC(\LamR)$ satisfying {\rm{(1.1.20)}}.
Any limit-point operator
$R^{\Lam_0}$ for $\left\{R^{\Lam_0}_{\Lam |\bx(\LamR)}\right\}$ is a positive-definite operator in
$\cH (\Lam_0)$ of trace $1$. Furthermore, let $\Lam_1\subset\Lam_0$ be a pair of cubes
and $R^{\Lam_1}$, $R^{\Lam_0}$ be a pair of limit-point RDMs  such that
$$R^{\Lam_1}=\lim\limits_{k\to +\infty}R^{\Lam_1}_{\Lam (k)|\bx(\Lam (k)^{(\tR)}}
\hbox{ and }R^{\Lam_0}=\lim\limits_{k\to +\infty}R^{\Lam_0}_{\Lam (k)|\bx(\Lam (k)^{(\tR)}}
\eqno (1.2.1)$$
for a sequence of cubes $\Lam (l)=[-L(l),L(l)]^{\times d}$ where $l=1,2,\ldots$, $L(l)\nearrow\infty$
and external CCs $\bx(\Lam (l)^{(\tR)})\in\cC(\Lam (l)^{(\tR)})$.
Then  $R^{\Lam_1}$ and $R^{\Lam_0}$ satisfy the compatibility property
$$R^{\Lam_1}=\tr_{\cH (\Lam_0\setminus\Lam_1)}R^{\Lam_0}. \qquad\Box\eqno (1.2.2)$$
}
\medskip

A direct consequence of Theorem 1.1 yields the construction of a limit-point infinite-volume
Gibbs state $\varphi$. For this purpose, it is enough to consider a countable family of cubes
$\Lam_0(l_0)=\left[-L_0l_0,L_0l_0\right]^{\times d}$ centered at the origin, of side-length
$2L_0l_0$, where $L_0\in (0,\infty )$
is fixed and $l_0=1,2,\ldots$ . By virtue of a diagonal process, we can ensure that,
given a family of external CCs  $\bx(\LamR)$, one can excerpt a sequence
$\Lam (l)\nearrow\bbR^d$ such that (a) $\forall$ natural $l_0$ $\;\exists\;$
the trace-norm limit
$$R^{\Lam_0(l_0)}=\lim\limits_{l\to +\infty}
R^{\Lam_0(l_0)}_{\Lam (l)|\bx(\Lam (l)^{(\tR)})}.\eqno (1.2.3)$$
and (b) for the limiting operators $R^{\Lam_0(l_0)}$ relation (1.2.1) is satisfied, with
$\Lam_1=\Lam_0(l_1)$ and $\Lam_0=\Lam_0(l_0)$
whenever $l_1<l_0$. This allows us to define an infinite-volume Gibbs state
$\varphi$ by setting
$$\varphi (A)=\lim\limits_{l\to\infty}\varphi_{\Lam (l)}(A)={\tr}_{\cH (\Lam_0(l_0))}
\big(AR^{\Lam_0(l_0)}\big),\;\;A\in\fB (\Lam_0(l_0)).\eqno (1.2.4)$$
More exactly, $\varphi$ is a state of the quasilocal C$^*$-algebra $\fB (\bbR^d)$
defined as the norm-closure of the inductive limit $\fB^0(\bbR^d)$:
$$\fB=\left( \fB^0(\bbR^d)\right)^-,\;\; \fB^0(\bbR^d)=
\operatornamewithlimits{\hbox{ind lim}}\limits_{\Lam\nearrow\bbR^d}
\fB (\Lam ).\eqno (1.2.5)$$
What is more, $\phi$ is defined by a family of finite-volume RDMs $R^{\Lam_0}$
acting in $\cH(\Lam_0)$, with $\Lam_0\subset\bbR^d$ being an arbitrary cube of the
form (1.1.13), and obeying the compatibility property (1.2.2).

\medskip
As we mentioned earlier, in two dimensions we prove the property of shift-invariance
of the limit-point Gibbs states $\varphi$. Note that $\forall$
cube $\Lam_0$ as in (1.1.13) and vector $s=\left({\ts}^1,\ldots ,{\ts}^d\right)\in\bbR^d$,
the Fock spaces $\cH (\Lam_0)$ and $\cH ({\tS}(s)\Lam_0)$ can be related via
mutually inverse shift isomorphisms:
$${\tU}^{\Lam_0}(s):\;\cH (\Lam_0)\to\cH ({\tS}(s)\Lam_0)\;\;\hbox{ and }\;\;
{\tU}^{\tS(s)\Lam_0}(-s):\; \cH ({\tS}(s)\Lam_0)\to\cH (\Lam_0).$$
Here ${\tS}({\ts})$ denotes the shift isometry $\bbR^d\to\bbR^d$:
$${\tS}(s):\;y\mapsto y+s,\;\;y\in\bbR^d,\eqno (1.2.6)$$
while ${\tS}(s)\Lam_0$ stands for the image of $\Lam_0$:
$$\begin{array}{l}{\tS}(s)\Lam_0
=\left[-L_0+{\tc}^1_0+{\ts}^1,{\ts}^1+{\tc}^1_0+L^0\right]\\
\quad\qquad\qquad\times\cdots\times
\left[-L_0+{\tc}^d_0+{\ts}^d,{\ts}^d+{\tc}^d_0+L^0\right].\end{array} \eqno (1.2.7)$$
The isomorphisms ${\tU}^{\Lam_0}(s)$ and
${\tU}^{\tS(s)\Lam_0}(-s)$ are defined as follows:
$$\begin{array}{c}\left({\tU}^{\Lam_0}(s)\phi_\unn\right) (\ux^\unn)
=\phi_\unn\left({\tS}(-s)\ux^\unn\right),\;\;
\ux^\unn\in\in\tS\Lam_0^\unn ,\; \phi_\unn\in\cH_\unn (\Lam_0),\\
\left({\tU}^{\tS(s)\Lam_0}(-s)\phi_\unn\right) (\ux^\unn)=
\phi_\unn ({\tS}(s)\ux^\unn),\;\;\ux^\unn\in
\Lam_0^\unn,\;\phi_\unn\in\cH_\unn (\Lam_0),\end{array}\eqno (1.2.8)$$
with $\unn=(n (1),\ldots n(q)),\;n(j)=0,1,\ldots $ .
\medskip

{\bf Theorem 1.2.} {\sl Let $d=2$ and $\beta\in (0, +\infty )$, $\uz\in (0,1)^q$.
Then any limit-point infinite-volume Gibbs state $\varphi$ is shift-invariant:
$\forall$ $s=(\ts^1,\ts^2)\in\bbR^{2}$
$$\varphi (A)=\varphi ({\tS}(s)A),\;\;A\in\fB(\bbR^{2}).\eqno (1.2.9)$$
Here ${\tS}(s)A$ stands for the shift of the argument $A$: if $A\in\fB (\Lam_0)$ where $\Lam_0$
is a square $[-L_0+\tc_1,\tc_1+L_0]\times [-L_0+\tc_2,\tc_2+L_0]$ then
$${\tS}(s)A ={\tU}^{\tS(s)\Lam_0}(-s)A\,{\tU}^{\Lam_0}(s)\in\fB({\tS}(s)\Lam_0).$$
In terms of the RDMs $R^{\Lam_0}$:
$$R^{\tS(s)\Lam_0} ={\tU}^{\Lam_0}(s)R^{\Lam_0}\,{\tU}^{\tS(s)\Lam_0}(-s).\qquad\Box\eqno (1.2.10)$$}

\subsection{The FK-representation for the RDMs}

Let us return to a general
value of dimension $d$. We will assume that $\beta\in (0, +\infty )$ and $\uz\in (0,1)^q$.
According to the featured realization of the Fock space $\cH (\Lam )$ as
${\rm L}_2(\cC (\Lam ))$ (see (1.1.19)), its elements are
identified as functions
$\phi_\Lam :\;\bx(\Lam )\in\cC (\Lam )\mapsto\phi_\Lam (\bx(\Lam ))\in{\bbC}$,
with
$$\diy\int_{\cC (\Lam )}
\left|\phi_\Lam (\bx(\Lam ) )\right|^2{\rd}\bx(\Lam )<\infty .
\eqno (1.3.1)$$
The space
$\cH (\Lam_0)$ is represented in a similar manner: here we will use  a short-hand
notation $\obx_0$ and $\oby_0$ instead of  $\bx(\Lam_0),\by (\Lam_0)\in\cC(\Lam_0)$. (When it is
convenient, $\obx_0$ and $\oby_0$  are understood as ordered arrays and identified with
$\ux^\unn$ and $\uy^\unn$, points from $\Lam_0^\unn$.)

The  first step in the proof of Theorems 1.1 is to reduce its assertions to statements
about  the integral kernels $F^{\Lam_0}_{\Lam |\bx(\LamR)}$ and
$F^{\Lam_0}$ which define the RDMs $R^{\Lam_0}_{\Lam |\bx(\LamR)}$
and their infinite-volume counterpart $R^{\Lam_0}$; we call these kernels RDMKs for short.
Indeed,  $R^{\Lam_0}_{\Lam |\bx(\LamR)}$ and $R^{\Lam_0}$ are integral
operators:
$$\begin{array}{c}\diy\left(R^{\Lam_0}_{\Lam |\bx(\LamR)}\phi_\Lam\right) (\obx_0)
=\int_{{\cC}(\Lam )} F^{\Lam_0}_{\Lam |\bx(\LamR)}(\obx_0,\oby_0)
\phi_\Lam (\oby_0){\rd}\oby_0\\
\diy\left(R^{\Lam_0}\phi_\Lam\right) (\obx_0)=\int_{{\cC}(\Lam )}
F^{\Lam_0}(\obx_0,\oby_0)
\phi_\Lam (\oby_0){\rd}\oby_0.\end{array}\eqno (1.3.2)$$
The RDMKs $F^{\Lam_0}_{\Lam |\bx(\LamR)}(\obx_0,\oby_0)$ and $F^{\Lam_0}(\obx_0,\oby_0)$
are investigated through an FK representation. Properties of these kernels are listed
in Theorem 1.3 where we adopt a setting from Theorem 1.1. We refer in Theorem 1.3  to the
Hilbert--Schmidt (HS) metric generated by the norm $\left\|\rho \right\|_{\rm{HS}}$ where
$$\left\|\rho \right\|^2_{\rm{HS}}=\int_{{\cC}(\Lam_0)\times{\cC}(\Lam_0)}
\left|\rho (\obx_0,\oby_0)\right|^2\rd\obx_0\rd\oby_0.\eqno (1.3.3)$$
Here $(\obx_0,\oby_0)\mapsto \rho  (\obx_0,\oby_0)$ is an integral kernel (in general, complex)
representing an HS operator in $\cH (\Lam_0)$. (Equivalently,
$\rho\in\cH (\Lam_0)\otimes\cH (\Lam_0)$.)
\medskip

{\bf Theorem 1.3.} {\sl Any pair of cubes $\Lam_0\subset\Lam$ and a family of
CCs $\bx(\LamR)\in\cC (\LamR)$ obeying {\rm{(1.1.20) }},
the family of RDMKs $F^{\Lam_0}_{\Lam |\bx(\LamR )}(\obx_0,\oby_0)$
is compact in the HS metric. Any limit-point function
$$(\obx_0,\oby_0)\in{\cC}(\Lam_0)\times{\cC}(\Lam_0)\mapsto
F^{\Lam_0}(\obx_0,\oby_0)\eqno (1.3.4)$$
determines a positive-definite operator $R^{\Lam_0}$  in $\cH (\Lam_0)$ of trace $1$
(a limit-point RDM). Furthermore, let $\Lam_1\subset\Lam_0$ be a pair of cubes
and $F^{\Lam_1}$, $F^{\Lam_0}$ a pair of limit-point RDMKs  such that
$$F^{\Lam_0}=\lim\limits_{k\to +\infty}F^{\Lam_0}_{\Lam (l)|\bx(\Lam (l)^{\rm c})}
\eqno (1.3.5)$$
in $C^0\left({\cC}(\Lam_0)\times{\cC}(\Lam_0)\right)$ for a sequence of cubes
$\Lam (l)\nearrow\bbR^d$ and
external CCs $\obx\left(\Lam (l)^{(\tR)}\right)$. Then the corresponding
limit-point RDMs  $R^{\Lam_1}$ and $R^{\Lam_0}$ obey {\rm{(1.2.2)}}.}  $\qquad\Box$
\medskip

Theorem 1.1 is deduced from Theorem 1.3 with the help of Theorem 1.4 below. The latter
is a slight generalisation of Lemma 1.5 from \cite{KS1} (going back to Lemma 1 in \cite{Su1}).
\medskip

{\bf Theorem 1.4.} {\sl
 Let $\rho_m(\tx,\ty)$, $\tx,\ty\in\tM$, be a sequence of kernels defining positive-definite
operators $R_m$ of trace class and with trace $1$ in a Hilbert
space $L_2(\tM,\nu )$ where $\nu (\tM)<\infty$. Suppose that as $m\to\infty$,
$\rho_m(\tx,\ty)$ converge to a limit kernel $\rho (\tx,\ty)$ in the Hilbert--Schmidt (HS) norm:
$$\begin{array}{ll}\diy\left\|\rho_m-\rho \right\|^2_{\rm{HS}}
=\int_{\tM\times\tM}\big[\rho_m(\tx,\ty)-\rho (\tx,\ty)\big]^2\nu (\rd\tx)\nu (\rd\ty)
&\to 0,\end{array}\eqno (1.3.6)$$
and $\rho (\tx,\ty)$ defines a positive-definite trace-class operator $R$
of trace $1$. Then
$$\lim_{m\to\infty}\|R_m-R\|_{\rm{tr}}=0\eqno (1.3.7)$$
where $\|A\|_{\rm{tr}}={\rm{tr}}\big(AA^*\big)^{1/2}$. $\quad\Box$}

\medskip

The proof Theorem 1.4 repeats that of the aforementioned lemmas, and
for shortness we do not reproduce it here.

\medskip

Therefore we focus from now on upon the proof of Theorems 1.2 and 1.3. In fact,
we will establish similar facts for more general objects -- FK-DLR functionals.
As in \cite{SK}, we use the terms a (multi-type) path configuration (PC) and a
(multi-type) loop configuration (LC).
The concept of FK-DLR functionals is based in our context on a series of definitions from
Sects 2.${\bI}$ and 4.${\bI}$ related to PCs and LCs. We will not repeat here Definitions 2.1.1.${\bI}$--2.1.4.${\bI}$
but give the list of the relevant notation used below. As to Definitions 2.1.1.${\bI}$,
2.1.2.${\bI}$, the corresponding objects are grouped into pairs: items with
the symbol $\cW$ represent path spaces whereas items with the symbol $\bbP$
represent path measures:
\medskip

(i) $\ocW^{\,k\beta}(x,y) \leftrightarrow \rd{\ov{\bbP}}^{\,k\beta}_{x,y}(\oom )\,$.\medskip

(ii) $\ocW^{\,*}(x,y)=
\operatornamewithlimits{\cup}\limits_{k\geq 1}\ocW^{k\beta}(x,y) \leftrightarrow
\rd{\ov{\bbP}}^{\,*}_{x,y}(\oom^*)\,$. \medskip

(iii) $\cW^*(x)=\ocW^{\,*}(x,x) \leftrightarrow \rd{\bbP}^*_x(\om^*)
=\rd{\ov{\bbP}}^{\,*}_{x,x}(\om^*)\,$. \medskip

(iv) $\ocW^*(\ux (j),\uy (j))=
\operatornamewithlimits{\times}\limits_{1\leq i\leq n(j)}\ocW^*(x_{j,i},y_{j,i})$ \\
$\leftrightarrow
{\ov{\bbP}}^{\,*}_{\ux(j),\uy(j)}(\oOm^*(j))=\operatornamewithlimits{\times}\limits_{1\leq i\leq n(j)}
{\ov{\bbP}}^*_{x_{j,i},y_{j,i}}(\oom^*_{j,i})\,$ where $\oOm^*(j)=(\oom^*_{j,1},\ldots ,\oom^*_{j,n(j)})$.
\medskip

(v) $\ucW^*(\ux (j),\uy (j))=\operatornamewithlimits{\cup}\limits_{\pi_{n(j)}\in\fS_{n(j)}}\ocW^*(\ux (j),\pi_{n(j)}
\uy (j)) $\\
$\leftrightarrow \rd{\ubbP}^*_{\ux (j),\uy (j)}(\oUps^*(j))=\sum\limits_{\pi_{n(j)}\in\fS_{n(j)}}
\rd{\ov{\bbP}}^*_{\ux (j),\; \pi_{n(j)}\uy (j)}(\oOm^*(j))$. Here $\fS_{n(j)}$ is the permutation group on
$n(j)$ elements, and $\pi_{n(j)}\uy (j) =(y_{j,\pi_{n(j)}(1)},\ldots ,y_{j,\pi_{n(j)}(n(j))})$.
Symbol $\oUps^*(j)$ covers all type $j$ PCs $\oOm^*(j)\in \ocW^*(\ux (j),\pi_{n(j)}\uy (j))$ where $\ux (j)$, $\uy (j)$ are fixed and $\pi_{n(j)}\in \fS_{n(j)}$ varies.
\medskip

(vi) $\ucW^*(\ux^{\unn} ,\uy^{\unn} )=\operatornamewithlimits{\times}\limits_{1\leq j\leq q}\ucW^*(\ux (j),\uy (j)) $\\
$\leftrightarrow \rd{\ubbP}^*_{\;\ux^{\unn} ,\uy^{\unn} }(\obUps^*)=
\operatornamewithlimits{\times}\limits_{1\leq j\leq q}
\rd{\ubbP}^*_{\ux (j),\uy (j)}(\oUps^*(j))\,$.

\medskip
(vii) $\cW^*(\bx(j) )=\operatornamewithlimits{\times}\limits_{x\in\bx(j)}\cW^*(x)
\leftrightarrow
\rd{\bbP}^*_{\bx(j)}(\bOm^*(j))=\operatornamewithlimits{\times}\limits_{x\in\bx(j)}
 \rd{\bbP}^*_x(\om^*_x)\,$ where $\bOm^*(j)=\{\om^*_x,\,x\in\bx(j)\}$.
\medskip

(viii) $\cW^*(\obx)=\operatornamewithlimits{\times}\limits_{1\leq j\leq q}
\cW^*(\bx(j))\leftrightarrow
\rd{\bbP}^*_{\obx}(\bOm^*)=\operatornamewithlimits{\times}\limits_{1\leq j\leq q}
 \rd{\bbP}^*_{\bx(j)}(\bOm^*(j))\,$ where $\bOm^*=(\bOm^*(1),...,\bOm^*(q))$.
\medskip

(ix) $\cW^*(\Lam )=\operatornamewithlimits{\cup}\limits_{\obx\in{\cC} (\Lam )}
\cW^*(\obx )\leftrightarrow {\rd}\obx\times{\bbP}^*_{\obx }({\rd}\bOm^*_\Lam)
=:{\rd}\bOm^*_\Lam $. Here $\cW^*(\Lam )$ is the space of (finite) multi-type LCs
$\bOm^*_\Lam$ with the initial/end points in $\Lam$ (however, the loops
constituting $\bOm^*_\Lam$ do not need to stay in $\Lam$).
\medskip

(x) $\cW^*(\bbR^d)$: the set of countable multi-type LCs
$\bOm^*=\bOm^*_{\bbR^d}$
such that their initial/end point CCs $\obx =(\bx (1),\ldots ,\bx (q))$
have no accumulation points in $\bbR^d$. A similar meaning is assigned to the notation
$\cW^*(\Lamc)$ and $\bOm^*_{\Lamc}$.
\medskip

To recapitulate, we list once again most of frequently used symbols below:
$$\begin{array}{l}
\oOm^*(j)=(\oom^*(j,1),\ldots ,\oom^*(j,n(j))) \hbox{\; a type $j$ PC (ordered), with fixed}\\
\qquad\hbox{initial/end points},\\
\oUps^*(j)=(\oom^*(j,1),\ldots ,\oom^*(j,n(j))) \hbox{\; a type $j$ PC (ordered), with permuted}\\
\qquad \hbox{end points},\\
\obUps^*=(\oUps^*(1),\ldots ,\oUps^*(q)) \hbox{\; a multi-type PC (ordered),  with permuted end}\\
\qquad \hbox{points},\\
 \Om^*(j) \hbox{\; a type $j$ loop collection (unordered), with a fixed initial/end CC},\\
\bOm^*=(\Om^*(1),...,\Om^*(q))\hbox{\; a multi-type LC, with a fixed initial/end CC},\\
\Om^*_\Lam(j) \hbox{\; a finite type $j$ LC with a varying initial/end CC in $\Lam$,}\\
\bOm^*_\Lam =(\Om^*_\Lam(1),...,\Om^*_\Lam(q)) \hbox{\; a finite multi-type LC with a varying initial}\\
\qquad \hbox{/end CC in $\Lam$,}\\
\bOm^*_{\Lamc}=(\Om^*_{\Lamc}(1),...,\Om^*_{\Lamc}(q))\hbox{\; a countable multi-type
LC with a varying}\\
\qquad \hbox{initial/end point CC in $\Lamc$}.
\end{array}$$

As in \cite{SK}, we also use the term a $\ttt$-section (of a
path/loop, and of a PC/LC) and employ the notation
$$\{\oOm^{\,*}\}(j,\ttt ), \{\oUps^{\,*}\}(j,\ttt ), \{\obUps^{\,*}\}(\ttt ),
\{\Om^{\,*}\}(j, \ttt ), \{\bOm^{\,*}\}(\ttt ), \{\bOm^{\,*}_\Lam\}(\ttt ),
\{{\bOm^{\,*}}_{\Lamc} \}(\ttt ),$$
similarly to \cite{SK}, Sect 2.1. Next, for a concatenation of two or more
configurations (CC, PC and/or LC) we use the symbol $\vee$.

Furthermore, we need a host of (integral) energy functionals $h(\; \cdot \;)$ and $h(\; \cdot \; |\; \cdot \;)$, viz.,
$$h(\bOm^*), h(\bOm^*_\Lam), h({\bOm^* }|\bOm^*_{\Lamc}),
h(\bOm^*_\Lam | \bOm^*_{\Lamc}),$$
and their versions $h(\; \cdot \;||\bx(\LamR)$. Next, 'counting' functionals$^{3)}$\footnote{$^{3)}$
Recall, functionals $K$ and $L$ are related to the (aggregated) time-length multiplicities.} are needed, e.g.,
$$ K(\oUps^*_0(j)), K(\Om^*_0 (j)), K(\Om^*_\Lam (j)), L(\Om^*_0(j)), L(\Om^*_\Lam (j)),\;\; 1\leq j\leq q.$$
Also, indicator functionals $\alpha_\Lam$ and $\chi^{\Lam_0}$ will be
used$^{4)}$\footnote{$^{4)}$Recall, $\alpha_\Lam$ requires that the PC/LC in the argument does not
leave $\Lam$, whereas $\chi^{\Lam_0}$ prevents it from entering $\Lam_0$ at ceratin time points.}, e.g.,
$$\alpha_\Lam(\obOm^*), \alpha_\Lam(\bOm^*), \alpha_\Lam(\bOm^*_\Lam ).$$

The above functionals are also used with concatenated arguments, viz.,
$$\begin{array}{c}
h\left(\obUps^*_0\vee\bOm^*_{\Lam\setminus\Lam_0}\big|\bOm^*_{\Lamc}\right),\;\;
h\left(\bOm^*_0\vee\bOm^*_{\Lam\setminus\Lam_0}\big|
\bOm^*_{\Lamc}\right),\\ \; \\
\chi^{\Lam_0}\left(\obUps^*_0\vee\bOm^*_{\Lam\setminus\Lam_0}\vee\bOm^*_{\Lamc}\right),\;\;
\chi^{\Lam_0}\left(\bOm^*_0\vee\bOm^*_{\Lam\setminus\Lam_0}\vee\bOm^*_{\Lamc}\right)
\end{array}$$
where $\obUps^*_0\in \ucW^*(\obx_0,\oby_0)$ and
$\bOm^*_0=\bOm^*_{\Lam_0}\in \cW^*(\Lam_0)$, see (1.3.8), (1.3.9). Cf.
Eqns (2.1.1.${\bfI}$)--(2.1.14.${\bfI}$). We also use the (standard) representation of the
partition function $\Xi\big[\Lam |\bx(\LamR )\big]$ (see Eqn (2.2.1.${\bfI}$)) and  RDMK
$F^{\Lam_0}_{\Lam |\bx(\LamR)}(\obx_0,\oby_0)$ (see
Eqns (2.2.2.${\bfI}$)--(2.2.5.${\bfI}$)). Next, Lemma 2.2.1.${\bfI}$, Definition 2.4.${\bfI}$,
Lemma 2.2.${\bfI}$  and Definitions 2.5.${\bfI}$--2.7.${\bfI}$ introduce the concepts
of the infinite-volume FK-DLR functionals, states and measures.

We employ the same notation $\fF$
(for infinite-volume FK-DLR functionals), $\fF_+$ (for infinite-volume FK-DLR states) and
$\fK$ (for infinite-volume FK-DLR probability measures) as in \cite{SK}. Recall,
$\mu\in\fK$ is a probability measure (PM) on $(\cW(\bbR^d),\fM (\bbR^d))$.
Here $\fM (\bbR^d)$ is the sigma-algebra generated by
cylinder events. In a probabilistic terminology, $\mu$ is a random marked point process
with marks from $\cW^*(0)$, the space of loops starting and ending up at the origin. Formally,
$\fM(\bbR^d)$ is the smallest sigma-algebra contained the `local` sigma-algebras $\fM (\Lam )$
$\forall$ cube $\Lam$. Cf. Sect. 3.${\bfI}$.

For any PM $\mu\in\fK$, the Ruelle bound (see Eqns (2.3.18.${\bI}$)--(2.3.20.${\bI}$)) holds
true, with $\orho =z$. Finally, the statements of Theorems 2.1.${\bfI}$ and 2.2.${\bfI}$
are carried through.

To summarize the FK-DLR representation: $\forall$ functional $\varphi\in\fF$,
the RDMK of an RDM $R^{\Lam_0}$ in $\Lam_0$ has the form: $\forall$ $\Lam\supseteq\Lam_0$
and $\obx_0,\oby_0\in\cC (\Lam)$ with $\sharp\,\bx_0(j)=\sharp\,\by_0(j)$,
$$\begin{array}{l}\diy F^{\Lam_0}(\obx_0,\oby_0) =\int_{\ucW^*(\obx_0,\oby_0)}\;
{\rd}{\ubbP}^*_{\obx_0,\oby_0}\;(\obUps^*_0)\\
\quad\diy\times\int_{\cW^*(\bbR^d)}
{\rd}\mu (\bOm^*_{\Lamc}){\mathbf 1}\Big(\bOm^*_{\Lamc}
\in\cW^{\,*}(\Lamc)\Big)\\
\qquad\qquad\diy\times\int_{\cW^*(\Lam \setminus\Lam_0)}{\rd}\bOm^*_{\Lam\setminus\Lam_0}
\chi^{\Lam_0}(\obUps^*_0\vee\bOm^*_{\Lam\setminus\Lam_0}\vee\bOm^*_{\Lamc})\qquad\qquad{}\\
\;\;\;\diy\times\prod\limits_{1\leq j\leq q}z_j^{K(\oUps^*_0(j))}\frac{z_j^{K(\Om^*_{\Lam\setminus\Lam_0}(j))}}
{L(\Om^*_{\Lam\setminus\Lam_0}(j))}\exp\left[-h\left(\obUps^*_0\vee\bOm^*_{\Lam\setminus\Lam_0}\big|
\bOm^*_{\Lamc}\right) \right].\end{array}\eqno (1.3.8)$$
Here $\mu$ is an FK-DLR measure (i.e., $\mu\in\fK$). This means that the restriction
$\mu\upharpoonright_{\fM (\Lam_0)}$ is determined by the Radon--Nikodym derivative
admitting the following representation: $\forall$ $\Lam\supseteq\Lam_0$ and
$\bOm^*_0\in\cW^*(\Lam_0)$,
$$\begin{array}{l}\diy\frac{\rd\mu\upharpoonright_{\fM (\Lam_0)}(\bOm^*_0)}{\rd\bOm^*_0}=
\int_{\cW^*(\bbR^d)}
{\rd}\mu (\bOm^*_{\Lamc}){\mathbf 1}\Big(\bOm^*_{\Lamc}
\in\cW^{\,*}(\Lamc)\Big)\\
\qquad\qquad\diy\times\int_{\cW^*(\Lam \setminus\Lam_0)}{\rd}\bOm^*_{\Lam\setminus\Lam_0}
\chi^{\Lam_0}(\bOm^*_0\vee\bOm^*_{\Lam\setminus\Lam_0}\vee\bOm^*_{\Lamc})\qquad\qquad{}\\
\quad\diy\times\prod\limits_{1\leq j\leq q}\frac{z_j^{K(\Om^*_0(j))}}{L(\Om^*_0(j))}
\frac{z_j^{K(\Om^*_{\Lam\setminus\Lam_0}(j))}}{L(\Om^*_{\Lam\setminus\Lam_0}(j))}
\exp\;\left[-h\left(\bOm^*_0\vee\bOm^*_{\Lam\setminus\Lam_0}\big|
\bOm^*_{\Lamc}\right) \right]. \end{array}\eqno (1.3.9)$$
(Recall, we use the notation
$\obUps^*_0=(\oUps^*_0(1),\ldots ,\oUps^*_0(q))\in \ucW^*(\obx_0,\oby_0)$
and $\bOm^*_0=(\Om^*_0(1),\ldots ,\Om^*_0(q))\in \cW^*(\Lam_0)$.)
Similar formulas hold true for RDMKs $F^{\Lam_0}_{\Lam |\bx(\LamR )}(\obx_0,\oby_0)$ and
the PMs $\mu^{\Lam_0}_{\Lam |\bx(\LamR )}$; see below.
\medskip

The rest of the paper is organized as follows. In Section 2 we analyse compactness properties
and prove Theorem 1.3.
Section 3 gives a brief sketch of the proof of Theorem 1.2.

\section{The compactness argument: proof of\\ Theorem 1.3}

\subsection{Uniform boundedness and HS convergence}

Let us fix a cube $\Lam_0$ of
side length $2L_0$ centered at $c=(\tc^1,\ldots ,\tc^d)$:
cf. Eqn (1.1.13).  The first step in the proof is to verify that, as $\Lam_0\subset\Lam$
and cube $\Lam\nearrow\bbR^d$,  the RDMK $F^{\Lam_0}_{\Lam |\bx(\LamR )}(\obx_0,\oby_0)$
(see (1.3.2)) form a compact family in $C^0(\cC (\Lam_0,\unn)\times\cC (\Lam_0,\unn))$ $\forall$
given $\unn $. (We want to stress that we work with pairs $(\obx_0,\oby_0)$ with
$\sharp\;\bx_0 (j)=\sharp\;\by_0 (j)$; otherwise $F^{\Lam_0}_{\Lam |\bx(\LamR )}(\obx_0,\oby_0)=0$.)
Clearly, the Cartesian product
$\cC (\Lam_0,\unn)\times\cC (\Lam_0,\unn)$ (the range of variable $(\obx_0,\oby_0)$ with given
$\sharp\,\bx_0 (j)=\sharp\,\by_0 (j)=n(j)$) is compact. As in \cite{KS1}, \cite{KS2}, \cite{KSY2},
it is convenient to employ the Ascoli--Arzela theorem, i.e., verify that, for a given $\unn$, the
functions $F^{\Lam_0}_{\Lam |\bx(\LamR )}(\obx_0,\oby_0)\upharpoonright_{\cC (\Lam_0,\unn)\times\cC
(\Lam_0,\unn)}$ are uniformly bounded and equi-continuous.
\medskip

Checking uniform boundedness for a fixed $\unn$ proceeds as follows.
$\forall$ $(\obx_0,\oby_0)\in \cC (\Lam_0,\unn)\times\cC (\Lam_0,\unn)$,
the RDMK  $F^{\Lam_0}_{\Lam |\bx(\LamR )}(\obx_0,\oby_0)$ satisfies,
$\forall$ $\Lam\subseteq\Lam^\prime\supseteq\Lam_0$,
$$\begin{array}{l}\diy F^{\Lam_0}_{\Lam |\bx(\LamR)}(\obx_0,\oby_0) =\int_{\ucW^*(\obx_0,\oby_0)}
{\rd}{\ubbP}^*_{\obx_0,\oby_0}(\obUps^*_0)\\
\qquad\diy\times
\int_{\cW^*(\Lam )}
{\rd}\mu (\bOm^*_{\Lam\setminus\Lam^\prime}){\mathbf 1}\Big(\bOm^*_{\Lam\setminus\Lam^\prime}
\in\cW^{\,*}(\Lam\setminus\Lam^\prime)\Big)\\
\qquad\diy\times\int_{\cW^*(\Lam^\prime\setminus\Lam_0)}{\rd}\bOm^*_{\Lam^\prime\setminus\Lam_0}
\chi^{\Lam_0}
\Big(\obUps^*_0\vee\bOm^*_{\Lam^\prime\setminus\Lam_0}\vee\bOm^*_{\Lam\setminus\Lam^\prime}\Big)\\
\quad\diy\times\alpha_{\Lam}\Big(\obUps^*_0\vee\bOm^*_{\Lam^\prime\setminus\Lam_0}\vee
\bOm^*_{\Lam\setminus\Lam^\prime}\Big)
\prod\limits_{1\leq j\leq q}z_j^{K(\oUps^*_0(j))}
\frac{z_j^{K(\Om^*_{\Lam^\prime\setminus\Lam_0}(j))}}{L(
\bOm^*_{\Lam^\prime\setminus\Lam_0}(j))}\\
\qquad\qquad\times\exp\;\Big[-h\left(\obUps^*_0\vee\bOm^*_{\Lam^\prime\setminus\Lam_0}\big|
\bOm^*_{\Lam\setminus\Lam^\prime}\vee\bx(\LamR)\right) \Big].\end{array}\eqno (2.1.1)$$
When $\Lam^\prime =\Lam_0$, this simplifies to
$$\begin{array}{l}\diy F^{\Lam_0}_{\Lam |\bx(\LamR)}(\obx_0,\oby_0) =\int_{\ucW^*(\obx_0,\oby_0)}
{\rd}{\ubbP}^*_{\obx_0,\oby_0}(\obUps^*_0)\prod\limits_{1\leq j\leq q}z_j^{K(\oUps^*_0(j))}\\
\diy\times\int_{\cW^*(\Lam )}
{\rd}\mu (\bOm^*_{\Lam\setminus\Lam_0}){\mathbf 1}\Big(\bOm^*_{\Lam\setminus\Lam_0}
\in\cW^{\,*}(\Lam\setminus\Lam_0)\Big)\chi^{\Lam_0}\Big(\obUps^*_0\vee\bOm^*_{\Lam\setminus\Lam_0}\Big)
\end{array}$$
$$\qquad\qquad\times
\alpha_{\Lam}(\obUps^*_0\vee\bOm^*_{\Lam\setminus\Lam_0})\exp\;\Big[-h\left(\obUps^*_0\big|
\bOm^*_{\Lam\setminus\Lam_0}\vee\bx(\LamR )\right)\Big]\eqno (2.1.2)$$
and leads to the bound
$$F^{\Lam_0}_{\Lam |\bx(\LamR)}(\obx_0,\oby_0)\leq Q^{\Lam_0}(\obx_0,\oby_0).\eqno(2.1.3)$$
where function $Q^{\Lam_0}(\obx_0,\oby_0)$ is specified below.

Namely, for $\obx_0=(\ux_0(1),\ldots ,\ux_0(q)), \oby_0=(\uy_0(1),\ldots ,\uy_0(q))\in
\operatornamewithlimits{\times}\limits_{1\leq j\leq q} \Lam_0^{n(j)}$, with
$\ux_0(j)=(x(j,1),\ldots ,x(j,n(j)))$ and $\uy_0(j)=(y(j,1),\ldots ,y(j,n(j)))$, the RHS in
(2.1.3) is given by
$$\diy Q^{\Lam_0}(\obx_0,\oby_0)=\int_{\ucW^*(\obx_0,\oby_0)}
{\rd}{\ubbP}^*_{\obx_0,\oby_0}(\obUps^*_0)
\prod\limits_{1\leq j\leq q}z_j^{K(\oUps^*_0(j))}\chi^{\Lam_0}\big(\oUps^*_0(j)\big)$$
$$\begin{array}{l}
\qquad\diy = \prod\limits_{1\leq j\leq q}\;\sum\limits_{\pi\in\fS_{n(j)}}\;
\prod\limits_{1\leq l\leq n(j)}\;\sum\limits_{k\geq 1}\;{z_j}^{k} \\
\qquad\qquad\qquad\diy\times\int_{\cW^{\beta k}(x(j,l),y( j,\pi l))}\bbP^{\beta k}_
{x(j,l),y(j,\pi l)}(\rd\om^*(j,l))
\chi^{\Lam_0}(\oom^*(j,l)).\end{array}\eqno (2.1.4)$$
Whenever $\sharp\,\ux (j)\neq\sharp\,\uy (j)$, the quantity $Q^{\Lam_0}(\obx_0,\oby_0)$ is set
to be $0$.
Recall, in (2.1.1), (2.1.2) and (2.1.4) we work with path configurations
$$\obUps^*_0=(\oUps^*_0(1),\ldots ,\oUps^*_0(q)), \;\; \oUps^*_0(j)=
(\oom^*(j,1),\ldots ,\oom^*(j,n(j)))$$
with permuted endpoints. Accordingly, $\fS_{n(j)}$ denotes the symmetric group on $n(j)$ elements; $\pi=\pi_{n(j)}$ is
a permutation of order $n(j)$ acting on `digits' $1,\ldots ,n(j)$. Cf. part (v) in the series of definitions
(i)--(x) in Section 1.3. The integral
$\diy\int_{\ucW^*(\obx_0,\oby_0)}$ in (2.1.4) (more precisely, the presence
of the indicator $\chi^{\Lam_0}(\oUps^*_0(j))$) yields the
$\bbP^{\beta k}_
{x(j,l),y(j,\pi_{n(j)}l)}$-probabilities that the paths $\oom^*(j,l)$, of a varying
time-length $\beta k$ ($=\beta k(j,l)$),
issued from point $x(j,l)$ and ending up at point $y(j,\pi l)$ do not enter cube $\Lam_0$
at times $\beta$, $2\beta$, $\ldots$, $\beta (k(j,l)-1)$. Formally,
$$\begin{array}{l}\diy\prod\limits_{1\leq l\leq n(j)}
\int_{\ocW^{\beta k}(x(j,l),y(j,\pi l))}\bbP^{\beta k}_{x(j,l),y(j,\pi l)}
(\rd\oom^*(j,l))\chi^{\Lam_0}(\oom^*(j,l))\\
\qquad\diy=\prod\limits_{1\leq l\leq n(j)}\bbP^{\beta k}_{x(j,l),y(j,\pi l)}
\Big(\oom^*(m\beta)\not\in\Lam_0\;\forall\;m=1,\ldots,k-1 \Big).\end{array}$$

For the future proof of the HS compactness we need to check that
$$\begin{array}{r}\diy\sum\limits_{\unn \;\geq \;\underline{0}}\;\frac{1}{(\unn!)^2}
\int\limits_{(\Lam_0)^{\times \unn}
\times (\Lam_0)^{\times \unn}}
\prod\limits_{1\leq j\leq q}\rd \bx_0(j)\rd \by_0(j)\Big[Q^{\Lam_0}(\obx_0,\oby_0)\Big]^2
\qquad{}\\ \diy =\int_{\cC (\Lam_0)\times\cC (\Lam_0)}\rd\obx_0\rd\oby_0\Big[Q^{\Lam_0}
(\obx_0,\oby_0)
\Big]^2<\infty\end{array}\eqno(2.1.5)$$
where $\unn! = \diy \prod\limits_{1\leq j \leq q}n(j)!$, and
$$\rd\ux_0(j)=\diy
\prod\limits_{1\leq l\leq n(j)}\rd x(j,l), \;\;
\rd\uy_0(j)=\diy\prod\limits_{1\leq j\leq (j)}\rd y,$$
$$\rd\obx_0=\prod\limits_{1\leq j \leq q}\frac{1}{\sharp\,\ux_0(i)!}\rd\ux_0 (j), \;\;
\rd\oby_0=\prod\limits_{1\leq j \leq q}\frac{1}{\sharp\,\ux_0(i)!}\rd\uy_0 (j). $$

To this end, we estimate first the integral in $\rd\by_0$, \\
$\diy\int_{\cC (\Lam_0,\unn)}\rd\oby_0\Big[Q^{\Lam_0}(\obx_0,\oby_0)\Big]^2$, for
$\sharp\,\bx_0(j)=\sharp\,\by_0(j)=n(j)$, does not exceed
$$\begin{array}{l}\diy
\prod\limits_{1\leq j\leq q}\;\sum\limits_{\pi\in\fS_{n(j)}}\;
\prod\limits_{1\leq l\leq n}\;\sum\limits_{k\geq 1}\;
\int_{\cW^{\beta k}(x(j,l),x(j, \pi l))}
\sum\limits_{1\leq m<k} {z_j}^k \\
\times\bbP^{\beta k}_{x(j,l),x(j,\pi l)}\Big(\oom^*(m\beta)\in\Lam_0,\;
\oom^*(m^\prime\beta)\not\in\Lam_0,1\leq m^\prime<k,m^\prime\neq m\Big) \\
\;\;\diy =\prod\limits_{1\leq j\leq q}\;\sum\limits_{\pi\in\fS_{n(j)}}\;
\prod\limits_{1\leq l\leq n(j)}\;\sum\limits_{k\geq 1}\;z_j^k\;
\bbP^{\beta k}_{x(j,l),x(j,\pi l)}\Big(\oom^*_{j,l}(m\beta)\in\Lam_0\end{array}$$
$$\qquad\qquad\qquad\qquad\qquad\diy\hbox{just for one value}\;m\in\{1,\ldots ,k-1\}\Big).
\eqno (2.1.6)$$

The next step in the proof of (2.1.5) is to decompose the permutation $\pi_{n(j)}$ into the product of cycles:
$\pi_{n(j)}=\gam_{1}\cdots\gam_s$, a cycle $\gam_i$ having length $n_i$ where
$n_1+\ldots +n_s=n(j)$ and starting at digit $t_i$ (say). Next, we take into account such
a decomposition, and for each cycle $\gam_i$ merge the paths
$x_0(t_i)\to x_0(\gam_it_i)$,  $x_0(\gam_it_i)\to x_0(\gam_i^2t_i)$, $\ldots$,
$x_0(\gam_i^{n_i-1}t_i)\to x_0(t_i)$
into a loop with the identical initial and endpoint $x_0(t_i)$ lying within $\Lam_0$.
In addition, each among the above
paths contains precisely one intermediate time point of the form $\beta m$, where $m$
is a positive integer
such that the path at this point lies in $\Lam_0$. It is not hard to see that for the
emerging loop $\oom^*$, of the
time-length $\beta M$ (say), the total number of time-points $\beta m$ such that
$m$ is a positive integer, $1\leq m<M$
and $\oom^*(\beta m)\in\Lam_0$ is always odd. So,
$$\begin{array}{l}\diy\int_{\cC (\Lam_0)\times\cC (\Lam_0)}\rd\obx_0\rd\oby_0
\Big[Q^{\Lam_0}(\obx_0,\oby_0)\Big]^2\leq \prod\limits_{1\leq j\leq q}
\sum\limits_{s\geq 0}\frac{1}{s!}\\
\quad\diy\times\Bigg[\sum\limits_{M\geq 2}{z_j}^M
\int_{\Lam_0}\rd x
\int_{\cW^{\beta M}(x,x)}\bbP^{\beta M}_{x,x}(\rd\oom^*){\mathbf 1}\Big(\oom^*(m\beta )
\in\Lam_0\\
\qquad\hbox{ for an odd number of values }m\in\{
1,\ldots,M\} \Big)
\Bigg]^s\\
\qquad\qquad\diy\leq\prod\limits_{1\leq j\leq q}\;\sum\limits_{s\geq 0}\frac{1}{s!}
\left[\upsilon (\Lam_0)\sum\limits_{M\geq 1}{z_j}^M
\right]^s<\infty\end{array}\eqno (2.1.7)$$
where $\upsilon (\Lam_0)$ stands for the Euclidean volume of cube $\Lam_0$.
\medskip

A similar argument remains valid for the limiting RDMK $F^{\Lam_0}(\obx_0,\oby_0)$,
beginning with the representation
$$\begin{array}{l}\diy F^{\Lam_0}(\obx_0,\oby_0) =\int_{\ucW^*(\obx_0,\oby_0)}
{\rd}{\ubbP}^*_{\obx_0,\oby_0}(\obUps^*_0)\prod\limits_{1\leq j\leq q}z_j^{K(\oUps^*_0(j))}\\
\qquad\quad\diy\times\int_{\cW^*(\bbR^d)}{\rd}\mu
(\bOm^*_{\bbR^d\setminus\Lam_0}){\mathbf 1}\Big(\bOm^*_{\bbR^d\setminus\Lam_0}
\in\cW^{\,*}(\bbR^d\setminus\Lam_0)\Big)\\
\qquad\times\chi^{\Lam_0}\Big(\obUps^*_0\vee\bOm^*_{\bbR^d\setminus\Lam_0}\Big)
\exp\Big[-h\left(\obUps^*_0\;\big|\;
\bOm^*_{\bbR^d\setminus\Lam_0}\right) \Big];\end{array}\eqno(2.1.8)$$
this again leads to the bound
$$F^{\Lam_0}(\obx_0,\oby_0)\leq Q^{\Lam_0}(\obx_0,\oby_0)\eqno(2.1.9)
$$
similar to (2.1.3).

Accordingly, we can write:
$$\Big[F^{\Lam_0}_{\Lam |\bx(\LamR)}(\obx_0,\oby_0)-F^{\Lam_0}(\obx_0,\oby_0)\Big]^2
\leq 4\Big[Q^{\Lam_0}(\obx_0,\oby_0)\Big]^2.\eqno (2.1.10)$$
\medskip

Let us outline the argument of compactness in the HS norm. After checking that
the family of the RDMKs $F^{\Lam_0}_{\Lam |\bx(\LamR)}(\obx_0,\oby_0)$ satisfies, for given
$\unn$ and $\Lam_0$, the assumptions of the Ascoli--Arzela theorem, we guaranty compactness in
$C^0(\cC (\Lam_0,\unn)\times\cC (\Lam_0,\unn))$. Hence, $\forall$ $\Lam_0$ and $\unn$, we
can extract a sequence $\{\Lam (s),\bx((\Lam (s))^{(\tR )}\}$ along which we have a
convergence
$$F^{\Lam_0}_{\Lam |\bx(\LamR)}(\obx_0,\oby_0)\to F^{\Lam_0}(\obx_0,\oby_0)$$
as $s\to\infty$ uniform in $(\obx_0,\oby_0)\in\cC (\Lam_0,\unn)\times\cC (\Lam_0,\unn)$.
By invoking
the diagonal process, we obtain a sequence $\{\Lam (s),\bx((\Lam (s))^{(\tR )}\}$
along which we have convergence for
a given $\Lam_0$ but $\forall$ $\unn$. Next, by using the Lebesgue dominated convergence
theorem,
we get from (2.1.5) and (2.1.10) that along our sequence,
$$\int_{\cC (\Lam_0)\times\cC (\Lam_0)}\Big[F^{\Lam_0}_{\Lam (s)|\bx ((\Lam (s))^{(\tR )}}
(\obx_0,\oby_0)-F^{\Lam_0}(\obx_0,\oby_0)\Big]^2\to 0.\eqno (2.1.11)$$
Then Theorem 1.4 implies that the RDMs $R^{\Lam_0}_{\Lam |\bx(\LamR)}$ converge to $R^{\Lam_0}$
in the trace norm. Finally, by inspecting a countable family of cubes $\Lam_0$, we get convergence
for all given $\Lam_0$, i.e., the compactness of states.

\subsection{Equicontinuity}

To verify the equi-continuity property of RDMKs
$F^{\Lam_0}_{\Lam |\bx(\LamR )}(\obx_0,\oby_0)$,
we have to check uniform bounds upon the gradients
$$\nabla_xF^{\Lam_0}_{\Lam |\bx(\LamR )}(\obx_0,\oby_0)\;\hbox{ and }\;
\nabla_yF^{\Lam_0}_{\Lam |\bx(\LamR )}(\obx_0,\oby_0).$$
Here $x=x(j,l)$ is one of the points in $\ux_0(j)$ and $y=y(j,\pi l)$ one of the points in $\uy_0(j)$,
$1\leq l\leq n(j)$, $1\leq j\leq q$.
Both cases are treated in a similar fashion; for definiteness, we consider
gradients  $\nabla_yF^{\Lam_0}_{\Lam |\bx(\LamR )}(\obx_0,\oby_0)$.

It can be seen from representation (2.1.2), (2.1.8) that there are two
contributions into the gradient. The first contribution comes from varying the
functional $\exp\left[
-h\Big(\obUps^*_{\,0}\vee\bOm^*_{\Lam\setminus\Lam_0}\big|\bx(\LamR )\Big)\right]$.
The second one emerges from varying the measure
${\obbP}^*_{\;\obx_0,\oby_0}$. (We
are interested only in variations related to a chosen point $y$.) Symbolically,
$$\begin{array}{l}\diy \nabla_yF^{\Lam_0}_{\Lam |\bx(\LamR )}(\obx_0,\oby_0)\\
\quad\diy =\int_{\cW^*(\Lam )}{\rd}\mu_{\Lam |\bx(\LamR)}
(\bOm^*_{\Lam\setminus\Lam_0}) \chi^{\Lam_0}(\bOm^*_{\Lam\setminus\Lam_0})
{\mathbf 1}\Big(\bOm^*_{\Lam\setminus\Lam_0}\in\cW^{\,*}(\Lam\setminus\Lam_0)\Big)\\
\quad\diy\times\Bigg\{
\int_{\ucW^*(\obx_0,\oby_0)}{\rd}{\ubbP}^*_{\obx_0,\oby_0}(\obUps^*_0)
\nabla_y\exp\left[-h\left(\obUps^*_0\big|\;
\bOm^*_{\Lam\setminus\Lam_0}\vee\bx(\Lam^{\tR})\right) \right]\\
\diy +\left(\nabla_y\int_{\ucW^*(\bx_0,\by_0)}
{\rd}{\ubbP}^*_{\bx_0,\by_0}(\obUps^*_0)\right)\exp\left[-h\left(\obUps^*_0\big|\;
\bOm^*_{\Lam\setminus\Lam_0}\vee\bx(\Lam^{\tR})\right) \right]\Bigg\}\end{array}$$
$$\qquad\quad
\diy\times\prod\limits_{1\leq j\leq q}z_j^{K(\oUps^*_0(j))}\frac{z_j^{K(\Om_{\Lam\setminus\Lam_0}(j))}}{
L(\Om_{\Lam\setminus\Lam_0}(j))}
\chi^{\Lam_0}\big(\obUps^*_0\big)
\alpha_\Lam\Big(\obUps^*_0\vee\bOm^*_{\Lam\setminus\Lam_0}\Big).
\eqno (2.2.1)$$

Let us analyze the parts involving the gradient. As before,
write $\obx_0=(\ux_0(1),\ldots ,\ux_0(q))$, $\oby_0=(\uy_0(1),\ldots ,\uy_0(q))$ and
$\obUps^*_0=(\oUps^*_0(1),\ldots ,\oUps^*_0(q))$. Here
$$\ux_0(j)=(x(j,1),\ldots ,x(j,n(j)),\;\uy_0(j)=(y(j,1),\ldots ,y(j,n(j))$$
and $\oUps_0(j)$ is a type $j$ PC formed by paths $\om (j,l)=\oom^*_{j,l}$,
of varying time-lengths $k=k(j,l)$ and with permuted end-points:
$\oUps_0(j)=(\oom^*_{j,1},\ldots ,\oom^*_{j,n(j)})$.

We have to focus on the following expression:
$$\begin{array}{l}
\diy \nabla_y\int_{\ucW^*(\obx_0,\oby_0)}
\rd{\ubbP}^*_{\;\bx_0,\by_0}(\obUps^*_0)\exp\;\left[-h\Big(\obUps^*_0\big|
\bOm^*_{\Lam\setminus\Lam_0}\vee\bx(\LamR )\Big)\right]\\ \;\\
\diy = \prod\limits_{1\leq j\leq q}\sum\limits_{\pi\in\fS_{n(j)}}\;
\nabla_y\prod\limits_{1\leq l\leq n(j)}\;
\sum\limits_{k\geq 1}\,z_j^k\int_{\cW^{\beta k}(x(j,l),x(j,l))}
{\bbP}^{\beta k}_{x(j,l),x(j,l)}({\rd}\om^*_{j,l})\end{array}$$
$$\begin{array}{l}\diy\times\exp\;\left[-h\Big(\bOm^*_0+\obZ_0^*\big|
\bOm^*_{\Lam\setminus\Lam_0}\vee\bx(\LamR )\Big)\right]\\
\qquad\qquad\qquad\diy \times
\exp\,
\left\{- \diy\frac{|x(j,l)-y(j, \pi l)|^2_{\rm{Eu}}}{2k\beta}\;\right\}.\end{array}\eqno (2.2.2)$$
(The indicators $\chi^{\Lam_0}$ and $\alpha_\Lam$ do not contribute and are omitted.)
Here $\obZ_0^*$ is a collection of straight paths: $\obZ_0^*=(\orZ^*(1),\ldots ,\orZ^*(q))$, with
$\orZ^*(j)=(\zeta^*_{j,1},\ldots, \zeta^*_{j,n(j)})$ where each $\zeta^*_{j,l}$ is a linear function
$$\zeta^*_{j,l}:\ttt\in\big[0,k\beta\big]\mapsto\frac{\ttt}{k \beta}(y(j,\pi l)-x(j,l)),
\;\;1\leq j\leq q,\;\; 1\leq l\leq n(j).\eqno (2.2.3)$$
Observe that the argument $\bOm^*_0$ in (2.2.2) represents a collection of loops
$\om^*(j,l)=\om^*_{j.l}$  beginning and ending at coinciding points $x(j,l)$.

Of course, the gradient will only affect the expression
$$\begin{array}{l}\diy\exp\,
\bigg[-\frac{|x(j,l)-y(j,\pi l)|^2_{\rm{Eu}}}{2k\beta}\\
\qquad -h\Big(\oom^*_{j,l}+\zeta^*_{j,l}\;\big|\;
\big[\{\bOm^*_0+\obZ^*_0\}\setminus\{\oom^*_{j,l}+\zeta^*_{j,l}\}\big]\vee\bOm^*_{\Lam\setminus\Lam_0}
\vee\bx(\LamR )\Big)\bigg]\end{array}$$
where $\pi\in\fS_{n(j)}$ and $y(j,\pi l)=y$. The subscript\; Eu\; stresses that we work with the Euclidean
norm/distance.

The first aforementioned contribution to the gradient emerges when we
differentiate the term
$$\exp\;\left[-h\Big(\oom^*_{j,l}+\zeta^*_{j,l}\;\big|\;
\big[\{\bOm^*_0+\obZ^*_0\}\setminus\{\oom^*_{j,l}+\zeta^*_{j,l}\}\big]\vee\bOm^*_{\Lam\setminus\Lam_0}
\vee\bx(\LamR )\Big)\right];$$
this contribution is more difficult to estimate.
The second comes from differentiating the term
$$\exp\,\Big[- |x(j,l)-y(j,\pi l)|^2_{\rm{Eu}}\big/(2k\beta )\;\Big].$$
It is easier to assess, and we refer the reader to \cite{SK} for a detailed argument about it.

Thus, we focus on the first contribution and write the corresponding
expression down: for $y=y(j,\pi l)$, and with $k=k(j,l)$,
$$\begin{array}{l}\diy\nabla_y
\exp\left[-h\Big(\om^*_{j,l}+\zeta^*_{j,l}\big|
\big[\{\bOm^*_0+\obZ^*_0\}\setminus\{\om^*_{j,l}+\zeta^*_{j,l}\}\big]\vee\bOm^*_{\Lam\setminus\Lam_0}
\vee\bx(\LamR )\Big)\right]\\
\;\;=-\exp\left[-h\Big(\om^*_{j,l}+\zeta^*_{j,l}\big|
\big[\{\bOm^*_0+\obZ^*_0\}\setminus\{\om^*_{j,l}+\zeta^*_{j,l}\}\big]\vee\bOm^*_{\Lam\setminus\Lam_0}
\vee\bx(\LamR )\Big)\right]\end{array}$$
$$\begin{array}{l}
\qquad\diy\times\int_0^{\beta}\rd\ttt \sum\limits_{1\leq m<k(j,l)}\Bigg\{\sum\limits_{\substack{1\leq m^\prime<k(j,l)\\ m^\prime\neq m}}
\nabla_yV_{j,j}\Big(|\om^*_{j,l}(\ttt +m\beta )\\
\qquad\qquad\diy +\zeta^*_{j,l}(\ttt+m\beta) -\om^*_{j,l}(\ttt +m^\prime\beta)-\zeta^*_{j,l}(\ttt +m^\prime\beta)|\Big)\\
\qquad\diy +\sum\limits_{\substack{1\leq j^\prime\leq q\\ j^\prime\neq j}}\;\sum_{1\leq l^\prime\leq n_{j^\prime}}
\;\sum_{1\leq m^\prime<k(j^\prime ,l^\prime)}
\nabla_yV_{j,j^\prime}\Big(|\om^*_{j,l}(\ttt +m\beta )\end{array}$$
$$\begin{array}{l}\qquad\qquad\diy +\zeta^*_{j,l}(\ttt+m\beta) -\om^*_{j^\prime,l^\prime}(\ttt +m^\prime\beta)-\zeta^*_{j^\prime ,l^\prime}(\ttt +m^\prime\beta)|\Big)\\
\qquad\diy +\sum\limits_{1\leq j^\prime\leq q}\Bigg[\sum\limits_{\om^*\in\Om^*_{\Lam\setminus\Lam_0}(j^\prime )}
\sum\limits_{1\leq m^\prime<k(\om^*)}\nabla_yV_{j,j^\prime}\Big(|\om^*_{j,l}(\ttt +m\beta )\\
\qquad\qquad\qquad\qquad +\zeta^*_{j,l}(\ttt+m\beta)
-\om^*(\ttt +m^\prime\beta)|\Big)\\
\qquad\diy +\sum\limits_{\ox\in\bx(\LamR,j^\prime )}
\nabla_yV_{j,j^\prime}\Big(|\om^*_{j,l}(\ttt +m\beta )+\zeta^*_{j,l}(\ttt+m\beta)-\ox|\Big)\Bigg]\Bigg\}\,.
\end{array}\eqno (2.2.4)$$
The initial observation is that the two first sums,  in the RHS of Eqn (2.2.4),
$\diy\sum\limits_{1\leq m<k(j,l)}\;\sum\limits_{\substack{1\leq m^\prime<k(j,l)\\ m^\prime\neq m}}$
and $\diy\sum\limits_{1\leq m<k(j,l)}\;\sum\limits_{\substack{1\leq j^\prime\leq q\\ j^\prime\neq j}}\;\sum_{1\leq l^\prime <n(j^\prime )}
\;\sum_{1\leq m^\prime<k(j^\prime ,l^\prime )}$, can be controlled uniformly in $\Lam$ in a straightforward manner.
Their input to (2.2.2) is bounded, respectively, by
$$\begin{array}{l}\diy 2{\ov V}^{\,(1)}\int_{\ucW^*(\obx_0,\oby_0)}
{\rd}{\ubbP}^*_{\obx_0,\oby_0}(\obUps^*_0)\prod\limits_{1\leq i\leq q}z_i^{K(\oUps^*_0(i))}\chi^{\Lam_0}\obUps^*_0\\
\qquad\quad\diy\times
\int_0^\beta\rd\ttt\sum\limits_{1\leq m<m^\prime<k(j,l)}{\mathbf 1}
\Big(|\om^*_{j,l}(\ttt +m\beta )+\zeta^*_{j,l}(\ttt+m\beta)\\
\qquad\qquad\qquad\diy-\om^*_{j,l}(\ttt +m^\prime\beta)-\zeta^*_{j,l}(\ttt +m^\prime\beta)|<\tR\Big)
\end{array}$$
and -- for $n(j)>1$ --
$$\begin{array}{l}\diy{\ov V}^{\,(1)}\int_{\ucW^*(\obx_0,\oby_0)}
{\rd}{\ubbP}^*_{\obx_0,\oby_0}(\obUps^*_0)\prod\limits_{1\leq i\leq q}z_i^{K(\oUps^*_0(i))}\chi^{\Lam_0}\obUps^*_0\\
\quad\diy\times1
\int_0^\beta\rd\ttt\diy\sum\limits_{1\leq m<k(j,l)}\;\sum\limits_{\substack{1\leq j^\prime\leq q\\ j^\prime\neq j}}\;
\sum_{1\leq l^\prime <n(j^\prime )}\;\sum_{1\leq m^\prime<k(j^\prime ,l^\prime )}{\mathbf 1}
\Big(|\om^*_{j,l}(\ttt +m\beta ) \\
\qquad\qquad\diy +\zeta^*_{j,l}(\ttt+m\beta)-\om^*_{j^\prime ,l^\prime}(\ttt +m^\prime\beta)-\zeta^*_{j^\prime ,l^\prime}
(\ttt +m^\prime\beta)|<\tR\Big)\,.\end{array}$$
(The fact that potentials $V_{j,j^\prime}$ may take the value $+\infty$
does not play a role in this bound.)

We only need to assess these expressions for given $\unn$ and $\Lam_0$.
Indeed, we upper-bound them by a `brute force':
$$\hbox{by }\;\;\frac{\beta}{2}\diy{\ov V}^{\,(1)}\Theta_2(z_j)
\prod\limits_{1\leq i\leq q} n(i)!\left(1+\Theta_0(z_i)\right)^{n(i)}:=A_1(\unn)$$
and
$$\begin{array}{l}\diy\hbox{by }\;\;\beta (q-1){\ov V}^{\,(1)}
\left(\sum\limits_{1\leq i^\prime\leq q}\Theta_1(z_{i^\prime})\right)\Theta_1(z_j)\\
\qquad\qquad\quad\diy \times\prod\limits_{1\leq i\leq q} n(i)!\; n(i)
\left(1+\Theta_0(z_i)\right)^{n(i)}:=A_2(\unn).\end{array}$$
(At this stage we did not use the fact that potentials $V_{j,j^\prime}$ have a finite radius.)
Here and below we use a host of quantities $\Theta_a(z)=\Theta_a(z,\beta )$:
$$\Theta_a(z)=\sum\limits_{k\geq 1}\frac{z^kk^a}{(2\pi\beta k)^{d/2}},\;\;a=-1,0,1,2.\eqno (2.2.5)$$

The third sum,
$\sum\limits_{1\leq m<k(j,l)}\;\sum\limits_{1\leq j^\prime\leq q}\;\sum\limits_{\om^*\in\Om^*_{\Lam\setminus\Lam_0}(j^\prime )}\;
\sum\limits_{1\leq m^\prime<k(\om^*)}$, in the RHS of (2.2.4)
involves loops $\om^*$ from the LC $\bOm^*_{\Lam\setminus\Lam_0}$. (Here $k(\om^*)$ stands for
the time-multiplicity of $\om^*$.) The contribution of this
sum into (2.2.1) is bounded from above in norm by
$$\begin{array}{l}\diy{\ov V}^{\,(1)}\int_{\ucW^*(\obx_0,\oby_0)}
{\rd}{\ubbP}^*_{\obx_0,\oby_0}(\obOm^*_0)\\
\;\;\diy\times\int_{\cW^*(\Lam )}{\rd}\mu_{\Lam |\bx(\LamR)}
(\bOm^*_{\Lam\setminus\Lam_0})
{\mathbf 1}\Big(\bOm^*_{\Lam\setminus\Lam_0}\in\cW^{\,*}(\Lam\setminus\Lam_0)\Big)\end{array}$$
$$\qquad\diy\times\chi^{\Lam_0}\Big(\obUps^*_0\vee\bOm^*_{\Lam\setminus\Lam_0}\Big)
\alpha_\Lam\Big(\obUps^*_0\vee\bOm^*_{\Lam\setminus\Lam_0}\Big)\eqno (2.2.6)$$
$$\begin{array}{l}\quad\diy\times\prod\limits_{1\leq i\leq q}z_i^{K(\oUps^*_0(i))}\frac{z_i^{K(\Om^*_0(i))}}{L(\Om^*_0(i))}
\exp\Big[-h\left(\obUps^*_0\;\big|\;
\bOm^*_{\Lam\setminus\Lam_0}\vee\bx(\Lam^{\tR})\right) \Big]\\
\diy\times\sum\limits_{\om^*\in\bOm^*_{\Lam\setminus\Lam_0}}
\int_0^\beta\rd\ttt\sum\limits_{\substack{1\leq m<k(j,l)\\1\leq m^\prime<k(\om^*)}}{\mathbf 1}
\Big(|\oom^*_{j,l}(\ttt +m\beta)-\om^*(\ttt +m^\prime\beta )|<\tR\Big).
\end{array}$$

By using the Campbell theorem, the Ruelle bound and the fact that\\ $h\left(\oUps^*_0+\obZ_0^*\;\big|\;
\bOm^*_{\Lam\setminus\Lam_0}\vee\bx(\Lam^{\tR})\right)\geq 0$ and omitting unused
indicators, the quantity (2.2.6) does not exceed
$$\diy{\ov V}^{\,(1)}\int_{\ucW^*(\obx_0,\oby_0)}
{\rd}{\ubbP}^*_{\obx_0,\oby_0}(\obUps^*_0)\prod\limits_{1\leq i\leq q}z_i^{K(\oUps^*_0(i))}$$
$$\qquad\diy\times\Bigg[\sum\limits_{1\leq i^\prime\leq q}
\int_{\bbR^d}{\rd}x\int_{\cW^*(x,x)}\rd\bbP^*_{x,x}(\om^*)
\;\frac{z_{i^\prime}^{k(\om^*)}}{k(\om^*)}\eqno (2.2.7)$$
$$\diy\times
\int_0^\beta\rd\ttt\sum\limits_{\substack{1\leq m<k(j,l)\\1\leq m^\prime<k(\om^*)}}{\mathbf 1}
\Big(|\oom^*_{j,l}(\ttt +m\beta)-\om^*(\ttt +m^\prime\beta )|<\tR\Big)\Bigg].$$
Observe, that the expression (2.2.7) does not depend upon $\Lam\supset\Lam_0$.

In turn,  (2.2.7) is less than or equal to
$$\diy\beta {\ov V}^{\,(1)}\int_{\ucW^*(\obx_0,\oby_0)}
{\rd}{\ubbP}^*_{\obx_0,\oby_0}(\obUps^*_0)\prod\limits_{1\leq i\leq q}z_i^{K(\oUps^*_0(i))}
\Bigg[\sum\limits_{1\leq i^\prime\leq q}\sum\limits_{k\geq 1}\;\frac{z_{i^\prime}^k}{k}\qquad$$
$$\qquad\times\diy\sum\limits_{1\leq m<k(j,l)}
\int_{\bbR^d}{\rd}x\;
\sum\limits_{1\leq m^\prime<k}\;\bbP^{\beta k}_{x,x}\Big(
\hbox{$\om^*\in\cW^{\beta k}(x,x):\;\om^*(\ttt +m^\prime\beta )$}\eqno (2.2.8)$$
$$\quad\hbox{lies within distance $\leq\tR\;$ from $\;\oom^*_{j,l}(\ttt +m\beta )$,
for some $\ttt\in [0,\beta ]$}\Big)\Bigg].$$

Next, by moving the starting/end points of both paths, $\om^*$
and  $\oom^*_{j,l}$, we obtain that (2.2.8) does not exceed
$$\begin{array}{l}\diy\beta {\ov V}^{\,(1)}\int_{\ucW^*(\obx_0,\oby_0)}
{\rd}{\ubbP}^*_{\obx_0,\oby_0}(\obUps^*_0)\prod\limits_{1\leq i\leq q}z_i^{K(\oUps^*_0(i))}k(j,l)\;
\sum\limits_{1\leq i^\prime\leq q}\sum\limits_{k\geq 1}\;\frac{z_{i^\prime}^k}{k}\\
\qquad\diy\times\int_{\bbR^d}{\rd}x\;\int_{\cW^{\beta k}(x,x)}\rd\bbP^{\beta k}_{x,x}(\om^*)
{\mathbf 1}\Big(
\hbox{$\om^*(\ttt )$ lies within distance}\\
\qquad\qquad\qquad\qquad\hbox{$\leq\tR\;$ from $\;\oom^*_{j,l}(\ttt )$,
for some $\ttt\in [0,\beta ]$}\Big).\end{array}\eqno (2.2.9)$$

To assess the RHS in (2.2.9), we use the requirement that the path
$\oom^*_{j,l}$ and the loop $\om^*$ must come close to each other on the time interval $[0,\beta ]$.
A necessary condition for this is that -- when ${\rm{dist}}_{\rm{Eu}}(x,\Lam_0)>\tR$ -- at least one
of them must travel at least a half of the distance
${\rm{dist}}_{\rm{Eu}}(x,\Lam_0)-\tR$ over the time interval $[0,\beta ]$. For a point $x\in\bbR^d$
with a large value of ${\rm{dist}}_{\rm{Eu}}(x,\Lam_0)$
it generates a sum of two small probabilities: one coming from $\bbP^{\beta k}_{x,x}$, the other
from $\obbP^{\,\beta k(j,l)}_{x(j,l),y(j,\pi l)}$. (Recall that  $y(j,\pi l)=y$, the varying point from $\oby_0$.)

Formally, we use Lemma 2.1:

\medskip

{\bf Lemma 2.1.} {\sl The following bounds hold true. {\rm{(i)}} $\forall$ $x\in \bbR^d$
and $a>0$,
$$
{\ubbP}^{\beta k}_{x,x}\Big(\sup\Big[{\rm{dist}}_{\rm{Eu}}(\omega^*(\ttt),\Lam_0)
\Big]:\;0\leq \ttt \leq \beta >a \Big)\leq c_0e^{-c_1 a^2}
$$
where $c_0=c_0(\beta)$ and $c_1=c_1(\beta)$ are finite positive constants.

{\rm{(ii)}} $\forall$ $x, y\in \Lam_0$
and $a>0$,
$$
{\ubbP}^{\beta k}_{x,y}\Big(\sup\Big[{\rm{dist}}_{\rm{Eu}}(\oom^*(\ttt),\Lam_0)
\Big]:\;0\leq \ttt \leq \beta >a \Big)\leq c_0e^{-c_1 a^2}
$$
where $c_0=c_0(\beta,\Lam_0)$ and $c_1=c_1(\beta,\Lam_0)$ are finite positive constants.
}
\medskip

{\it Proof of Lemma} 2.1. The starting point is the Skorohod formula for the Brownian bridge on the time
interval $[0,\beta ]$ in one dimension: given $a>0$ and  $\rx,\ry\in\bbR$ with $|\rx-\ry |<a$,
$$\begin{array}{l}\diy P^{\beta}_{\rx,\ry}\left(\om :\sup\Big[|\om (\ttt )-\rx|:\;0\leq\ttt\leq\beta\Big]>a\right)\\
\qquad\qquad\diy =\frac{1}{\sqrt{2\pi\beta}}\;\sum\limits_{l\in\bbZ:\;l\neq 0}\;(-1)^{l-1}
\exp\left [-\frac{1}{2\beta}(\ry-\rx-2la)^2\right].\end{array}\eqno (2.2.10)$$
(Here and later we use the notation $P^{\bullet}_{\bullet ,\bullet}$ for the (non-normalised) Wiener measure
of the bridge in one dimension.) Cf. \cite{S2}, Chapter 6, Sect. 27, Eqn (27.1) and below.
We convert it to the following equality:
$$\diy P^{\beta k}_{\rx,\ry}\left(\sup\Big[|\om (\ttt )-\rx|:0\leq\ttt\leq\beta\Big]>a\right)
=\frac{1}{2\pi\beta}\frac{1}{\sqrt{k-1}}\qquad\qquad\qquad$$
$$\qquad\qquad\quad\diy \times\;\int\rd\ru\Bigg\{{\mathbf 1}\big(|\rx -\ru |>a\big)
\exp\left [-\frac{(\ru-\rx)^2}{2\beta}-\frac{(\ru-\ry)^2}{2\beta (k-1)}\right]\eqno (2.2.11)$$
$$\quad\diy-{\mathbf 1}\big(|\rx -\ru |<a\big)
\sum\limits_{l\in\bbZ:\:l\neq 1}\exp\left[-\frac{(\ru-\rx -2la)^2}{2\beta}-\frac{(\ru-\ry)^2}{2\beta (k-1)}
\right]\Bigg\}\,.$$
(We agree that for $k=1$, (2.2.11) morphs back to (2.2.10).)

(i) Take  $\rx =\ry$. By the Cauchy--Schwarz inequality, the contribution of the integral
$\diy\int\rd\ru\;{\mathbf 1}\big(|\rx -\ru |>a\big)$ is
$$\leq\frac{1}{(4\pi^2\beta^2(k-1))^{1/4}}
\left(\frac{1}{\sqrt{\pi\beta}}\int\rd\ru\;{\mathbf 1}\big(|\rx -\ru |>a\big)
\exp\left [-\frac{(\ru-\rx)^2}{\beta}\right] \right)^{1/2}$$
$$=\frac{2}{(4\pi^2\beta^2(k-1))^{1/4}}\Phi\left(\frac{a}{2\sqrt\beta}\right)^{1/2}\hbox{ where }
\Phi (b)=\frac{1}{\sqrt{2\pi}}\int_b^{+\infty}e^{-\rv^2/2}\rd\rv.\eqno (2.2.12)$$

Next, consider the contribution of the integral $\diy\int\rd\ru\;{\mathbf 1}\big(|\rx -\ru |<a\big)$. When
$a>\ru -\rx >0$, we can write
$$\sum\limits_{l\in\bbZ:\:l\neq 1}(-1)^{l-1}\exp\left[-\frac{(\ru-\rx -2la)^2}{2\beta}-\frac{(\ru-\rx)^2}{2\beta (k-1)}
\right]\qquad\qquad$$
$$\leq \exp\left[-\frac{(\ru-\rx)^2}{2\beta (k-1)}\right]\Bigg\{\exp\left[-\frac{(\ru-\rx +2a)^2}{2\beta}\right]
\eqno (2.2.13)$$
$$\qquad\qquad-\exp\left[-\frac{(\ru-\rx +4a)^2}{2\beta}\right]
+\exp\left[-\frac{(\ru-\rx +6a)^2}{2\beta}\right] $$
$$+\exp\left[-\frac{(\ru-\rx -2a)^2}{2\beta}\right]
-\exp\left[-\frac{(\ru-\rx -4a)^2}{2\beta}\right]\qquad\qquad\qquad\qquad$$
$$+\exp\left[-\frac{(\ru-\rx -6a)^2}{2\beta}\right]\Bigg\}
\leq 6\exp\left[-\frac{(\ru-\rx)^2}{2\beta (k-1)}\right]\exp\left(-\frac{ a^2}{2\beta}\right).$$

A similar bound holds when $\ru<\rx$. Integrating in $\;\rd\ru\;$ yields a finite value, with the factor
$e^{-a^2/(2\beta)}$ in front.

Going back to (2.2.11) we can write
$$
\diy P^{\beta k}_{\rx,\ry}\left(\sup\Big[|\om (\ttt )-\rx|:0\leq\ttt\leq\beta\Big]>a\right)\leq
c_0\exp(c_1 a^2)
$$
where $c_0, c_1\in (0, +\infty)$ are constants depending upon $\beta$.
The rest of the argument completing the proof assertion (i) is standard and omitted.

The proof of statement (ii) is similar. $\qquad\blacksquare$
\medskip

By virtue of Lemma 2.1, we can
upper-bound the RHS of (2.2.9) by
$$\begin{array}{l}\diy2\beta {\ov V}^{\,(1)}
\left(\sum\limits_{1\leq i^\prime\leq q}\Theta_0(z_{i^\prime})\right)\Theta_1(z_j)
\prod\limits_{1\leq i\leq q} n(i)!\;n(i)
\left(1+\Theta_0(z_i)\right)^{n(i)}\end{array}\eqno (2.2.14)$$
$$\qquad\quad\diy\times\left\{c+
c_0\int_{\bbR^d}{\rd}x\exp\left[-{c_1}{\rm{dist}}_{\rm{Eu}}(x,\Lam_0)^{2}
\right]\right\}:=A_3(\unn ,\Lam_0).$$
Here $c\in (0,\infty )$, $c_0\in (0,\infty )$  and $c_1\in (0,\infty )$ are
constants.
\medskip

Let us now focus on the forth sum, $\sum\limits_{1\leq m<k(j,l)}\;
\sum\limits_{\substack{1\leq j^\prime\leq q\\ j^\prime\neq j}}\;\sum\limits_{\ox\in\bx(\LamR,j^\prime )}$, in
the RHS of (2.2.4). This sum contributes into (2.2.1) a quantity whose norm is
$$\leq{\ov V}^{\,(1)}\int_{\ucW^*(\obx_0,\oby_0)}
{\rd}{\ubbP}^*_{\obx_0,\oby_0}(\obUps^*_0)\prod\limits_{1\leq i\leq q}z_i^{K(\oUps^*_0(i))}
\chi^{\Lam_0}\obUps^*_0\alpha_\Lam\big(\obUps^*_0\big)$$
$$\begin{array}{l}\diy\;\;\diy\times\int_{\cW^*(\Lam )}{\rd}\mu_{\Lam |\bx(\LamR)}
(\bOm^*_{\Lam\setminus\Lam_0})
\chi^{\Lam_0}(\bOm^*_{\Lam\setminus\Lam_0})
{\mathbf 1}\Big(\bOm^*_{\Lam\setminus\Lam_0}\in\cW^{\,*}(\Lam\setminus\Lam_0)\Big)\\
\;\;\diy\times\prod\limits_{1\le i^\prime\le q}\frac{z_{i^\prime}^{K(\Om^*_{\Lam\setminus\Lam_0})}}{
L(\Om^*_{\Lam\setminus\Lam_0})}
\alpha_\Lam\Big(\bOm^*_{\Lam\setminus\Lam_0}\Big)
\exp\Big[-h\left(\obUps^*_0\;\big|\;
\bOm^*_{\Lam\setminus\Lam_0}\vee\bx(\Lam^{\tR})\right) \Big]\end{array}$$
$$\qquad\qquad\diy\times
\int_0^\beta\rd\ttt\sum\limits_{\substack{1\leq m<k(l,j)\\ \ox\in\bx (\LamR)}}{\mathbf 1}
\Big(|\oom^*_{j,l}(\ttt +m\beta)-\ox |<\tR\Big).
\eqno (2.2.15)$$

The middle integral in (2.2.15) is
$$\begin{array}{l}
\diy\int_{\cW^*(\Lam )}{\rd}\mu_{\Lam |\bx(\LamR)}(\bOm^*_{\Lam\setminus\Lam_0})
\chi^{\Lam_0}(\bOm^*_{\Lam\setminus\Lam_0})
{\mathbf 1}\Big(\bOm^*_{\Lam\setminus\Lam_0}\in\cW^{\,*}(\Lam\setminus\Lam_0)\Big)\\
\qquad\diy\times
\prod\limits_{1\le i^\prime\le q}\frac{z_{i^\prime}^{K(\Om^*_{\Lam\setminus\Lam_0}(i^\prime)}}{
L(\Om^*_{\Lam\setminus\Lam_0}(i^\prime ))}\alpha_\Lam (\bOm^*_{\Lam\setminus\Lam_0})
\exp\Big[-h\left(\obUps^*_0\;\big|\;
\bOm^*_{\Lam\setminus\Lam_0}\vee\bx(\Lam^{\tR})\right) \Big]\leq 1.
\end{array}$$
Indeed,  $\mu_{\Lam |\bx(\LamR)}$ is a probability distribution, the values $z_{i^\prime}\in (0,1)$,
functionals $K(\Om^*_{\Lam\setminus\Lam_0}(i^\prime)),L(\Om^*_{\Lam\setminus\Lam_0}(i^\prime))\geq 1$
and $h\left(\obUps^*_0\;\big|\;
\bOm^*_{\Lam\setminus\Lam_0}\vee\bx(\Lam^{\tR})\right)\geq 0$, and the rest are
indicators. Therefore, (2.2.15) does not exceed
$$\begin{array}{l}\diy{\ov V}^{\,(1)}\int_{\ucW^*(\obx_0,\oby_0)}
{\rd}{\ubbP}^*_{\obx_0,\oby_0}(\obUps^*_0)\prod\limits_{1\leq i\leq q}z_i^{K(\oUps^*_0(i))}
\chi^{\Lam_0}(\obUps^*_0)\\
\qquad\qquad\diy\times
\int_0^\beta\rd\ttt\sum\limits_{\substack{1\leq m<k(j,l)\\ \ox\in\bx(\LamR)}}{\mathbf 1}
\Big(|\oom^*_{j,l}(\ttt +m\beta)-\ox |<\tR\Big).
\end{array}\eqno (2.2.16)$$

To bound (2.2.16) from above, we use the following argument. The sum
$\diy\sum\limits_{\substack{1\leq m<k(j,l)\\ \ox\in\bx(\LamR)}}$ is not zero
only if the path $\oom^*_{j,l}$ reaches the `internal` annulus
$$\Lam_{(\tR )}=\left\{x\in\Lam :\;{\rm{dist}}_{\rm{Eu}}(x,\partial\Lam )\leq\tR\right\};$$
in this case the sum does not exceed $k(j,l)\sharp\,\bx(\LamR)$. The probability
that $\oom^*_{j,l}$ reaches $\Lam_{(\tR )}$ is
$$\le \diy\frac{1}{(2\pi\beta k(j,l))^{d/2}}\exp\,\left[-\frac{{\rm{dist}}_{\rm{Eu}}
(\Lam_0,\Lam_{(\tR )})^2}{2\beta k(j,l)}\right].$$
In turn, for $\Lam\supset\Lam_0$ we have that
$${\rm{dist}}_{\rm{Eu}}(\Lam_0,\Lam_{(\tR )})\geq L-\tR-L_0-{\rm{dist}}_{\rm{Eu}}(0,\Lam_0)$$
where ${\rm{dist}}_{\rm{Eu}}(0,\Lam_0)$
is the distance between $\Lam_0$ and the origin. Going back to the external
annulus $\LamR=\Lam_L^{(\tR )}$ (see (1.1.16)), the quantity (2.2.16) is
$$\diy\leq \beta{\ov V}^{\,(1)} \;\sharp\,\bx(\LamR)
\;\prod\limits_{1\leq i\leq q}n(i)!\left(1\vee
\sum\limits_{k\geq 1}\;\frac{z_i^k}{(\sqrt{2\pi\beta k})^d}\right)^{n(i)}$$
$$\diy\times\sum\limits_{k\geq 1}
\frac{z_j^kk}{(2\pi\beta k)^{d/2}}\exp\left[-\frac{
\big(L-\tR-L_0-{\rm{dist}}_{\rm{Eu}}(0,\Lam_0)\big)^2}{2\beta k}\right]$$
which in turn does not exceed
$$\beta{\ov V}^{\,(1)}\prod\limits_{1\leq i\leq q}n(i)!\left(1+\Theta_0(z_i)\right)^{n(i)}
 B(\rc):=A_4(\unn,\Lam_0).\eqno (2.2.17)$$
Here the quantity $B(\rc)$ has been introduced in Eqn (1.1.20), and the argument $\rc$ is specified as
$$\rc=(\tR+L_0+{\rm{dist}}_{\rm{Eu}}(0,\Lam_0))^2.\eqno (2.2.18)$$
\medskip

We see that the norm of the gradient vector represented by (2.2.1) is upper-bounded by
$$\begin{array}{r}\diy {\ov V}^{\,(1)}\Big[A_1(\unn,\Lam_0) +A_2(\unn,\Lam_0)
+A_3(\unn,\Lam_0)+ A_4(\unn,\Lam_0)\Big]\end{array}$$
which yields the equi-continuity property required.
Hence, the family of RDMKs $\left\{F^{\Lam_0}_{\Lam |\bx(\LamR )}\right\}$ is compact in
space $C^0(\cC (\Lam_0,\unn)\times\cC (\Lam_0,\unn))$. This closes the argument that
the set of Gibbs states $\varphi_{\Lam |\bx(\LamR)}$ is compact.

\subsection{Weak compactness of FK-DLR measures}

A version of the above argument
is applicable for proving that, for any given cube
$\Lam_0$, the probability measures (PMs)
$\mu^{\Lam_0}_{\Lam |\bx(\LamR )}$ on $\cW^*(\Lam_0)$ form a compact family as
$\Lam\nearrow\bbR^d$. According to the Prokhorov theorem,
it is enough to verify that  the family $\left\{\mu_{\Lam |\bx(\LamR )}^{\Lam_0}\right\}$
is tight.

The proof of tightness proceeds along steps (a)--(d); see below.
\medskip

(a) Let $\epsilon >0$ be given.
Then we can find $k^0=k^0(\epsilon ,\Lam_0)$ such that the value
$$\begin{array}{l}\mu^{\Lam_0}_{\Lam |\bx(\LamR )}\Big(\bOm^*_0=(\Om^*_0(1),\ldots ,
\Om^*_0(q))\in\cC(\Lam_0):\\
\qquad\qquad\qquad\max\big[K(\Om^*_0(j)):\;1\leq j\leq q\big]
\geq k^0\Big)\end{array}\eqno (2.3.1)$$
can be made as small as desired. In fact,
$$\begin{array}{l}\diy\mu^{\Lam_0}_{\Lam |\bx(\LamR )}\left(\bOm^*_0:\;
\max\big[K(\Om^*_0(j)):\;1\leq j\leq q\big]\geq k^0\right)\\
\;\;\diy =\int_{\cW^*(\Lam )}\rd\mu_{\Lam |\bx(\LamR )}(\bOm^*_{\Lam\setminus\Lam_0})
{\mathbf 1}(\bOm^*_{\Lam\setminus\Lam_0}\in\cW^*(\Lam\setminus\Lam_0))\\
\;\;\diy\times\int_{\cW^*(\Lam_0)}\rd\bOm^*_0\;
\prod\limits_{1\leq j\leq q}\frac{z_j^{K(\Om^*_0(j))}}{L(\Om^*_0(j))}\;{\mathbf 1}
(\max\big[K(\Om^*_0(j)):\;1\leq j\leq q\big]\geq k^0) \end{array}$$
$$\begin{array}{l}\;\;\diy\times\chi^{\Lam_0}(\bOm^*_0\vee
\bOm^*_{\Lam\setminus\Lam_0})\;\alpha_\Lam(\bOm^*_0\vee\bOm^*_{\Lam\setminus\Lam_0})
 \exp\Big[-h(\bOm^*_0\,\big|\,\bOm^*_{\Lam\setminus\Lam_0}\vee\bx(\LamR ))\Big]\\
 \quad\diy\leq\int_{\cW^*(\Lam_0)}\rd\bOm^*_0
\;\prod\limits_{1\leq j\leq q}\frac{z_j^{K(\Om^*_0(j))}}{L(\Om^*_0(j))}\;{\mathbf 1}
(\max\big[K(\Om^*_0(j)):\;1\leq j\leq q\big]\geq k^0)\end{array}$$
$$\diy =\prod\limits_{1\leq j\leq q}\exp\Big[\ups(\Lam_0)\left(1+\Theta_0(z_j)\right)\Big]\eqno (2.3.2)$$
$$\diy\times\sum\limits_{1\leq i\leq q}\sum\limits_{n\geq 0}\frac{\ups (\Lam_0)^n}{n!}
\prod\limits_{1\leq l\leq n}
\sum\limits_{k(l)\geq 1}\frac{z_i^{k(l)}}{k(l)(2\pi\beta k(l))^{d/2}}
{\mathbf 1}\left(\sum\limits_{1\leq l\leq n}k(l)\geq k^0\right).$$
Like before, $\ups (\Lam_0)$ stands here for the Euclidean volume of $\Lam_0$. For the
definition of $\Theta_0$ (and $\Theta_{-1}$ below),  see (2.2.5).

The sum $\sum\limits_{n\geq 0}$ in the RHS of (2.3.2) is divided into two: $\,\sum_1:=\sum\limits_{n>{\sqrt{k^0}}}\;$
and $\,\sum_2:=\sum\limits_{n\leq{\sqrt{k^0}}}$. The contribution of the former to the
last line in (2.3.2) is
$$\begin{array}{l}\diy\leq\sum\limits_{1\leq i\leq q}\sum\limits_{n>{\sqrt{k^0}}}
\frac{\upsilon (\Lam_0)^n}{n!}\Theta_{-1}(z_i)^n \end{array}\eqno (2.3.3)$$
which can be made arbitrarily small for large $k^0$. Next, in the latter at least
one $k(l)$ must satisfy $k(l)\geq k^0/n\geq{\sqrt{k^0}}$. So, the contribution
from $\sum_2$ to the last line in (2.3.2) does not exceed
$$\begin{array}{l}\diy\sum\limits_{1\leq i\leq q}
\left(\sum\limits_{k\geq{\sqrt{k^0}}}\frac{z_i^k}{k(2\pi\beta k)^{d/2}}\right)
\sum\limits_{n\leq{\sqrt{k^0}}}n\frac{\upsilon (\Lam_0)^n}{n!}\Theta_{-1}(z_i)^{n-1}
\end{array}\eqno (2.3.4)$$
which again is small for large $k^0$.

(b) The second step is the remark that the Radon-Nikodym derivative is bounded
uniformly in $\Lam$ and $\bx (\LamR)$ (since $z\in (0,1)$):
$$\begin{array}{l}\diy\frac{\rd\mu^{\Lam_0}_{\Lam |\bx(\LamR )}(\bOm^*_0)}
{\rd\bOm^*_0}\\
\quad\diy =\int_{\cW^*(\Lam )}\rd\mu_{\Lam |\bx(\LamR )}
(\bOm^*_{\Lam\setminus\Lam_0}){\mathbf 1}
(\bOm^*_{\Lam\setminus\Lam_0}\in\cW^*(\Lam\setminus\Lam_0))\\
\qquad\diy\times\chi^{\Lam_0}(\bOm^*_0\vee
\bOm^*_{\Lam\setminus\Lam_0})\;\prod\limits_{1\leq j\leq q}
\frac{z_j^{K(\Om^*_0(j))}}{L(\Om^*_0(j))}\;
\alpha_\Lam(\bOm^*_0\vee\bOm^*_{\Lam\setminus\Lam_0})\leq 1.
\end{array}\eqno (2.3.5)$$

(c) By virtue of property (b),  it suffices to prove that, for given
$\delta >0$ and positive integer $k^0$, there exists a compact set
$\cJ\subset\cC (\Lam_0)$ such that
$$\cJ\subset\cK(k^0):=\Big\{\bOm^*_0:\;\max\big[K(\Om^*_0(j))\big]
\leq k^0\Big\}\;\hbox{
and }\;\int_{\cC(\Lam_0)\setminus\cJ}\rd\bOm^*_0<\delta .\eqno (2.3.6)$$

As before, this is achieved with the help of the Ascoli--Arzela theorem,
connecting compactness with uniform boundedness and equi-continuity.
First, we guarantee the uniform boundedness by claiming that $\forall$
$\delta$ and $k_0$ there exists an $\ell^0\in (0,\infty )$ such that
$$\begin{array}{r}\diy\int_{\cK (k^0)}\rd\bOm^*_0\;{\mathbf 1}
\bigg(\operatornamewithlimits{\max}\limits_{\om^*\in\bOm^*_0}
\sup\,\Big[|\om^*(\ttt )-\om^*(0)|: 0\leq\ttt\leq\beta k(\om^*)\Big]
\geq \ell^0\bigg)\leq\frac{\delta}{2}.\end{array}\eqno (2.3.7)$$
This claim holds because on the set $\cK (k^0)$ the number of loops
$\om^*$ constituting the LC $\bOm^*_0$ and their time-multiplicities
$k(\om^*)$ do not exceed $k^0$.

(d) Finally, we need to verify the equi-continuity property.
But this fact holds true since the reference measure ${\rd}\bOm^*_0$ on
the set $\cK(k^0)$ is supported by LCs $\bOm^*_0$ such that all loops
$\om^*\in\bOm^*$ have a (global) continuity modulus not exceeding $\sqrt{2k^0\beta\epsilon\ln\,(1/\epsilon)}$.

This completes the proof of compactness for PMs
$\mu_{\Lam |\bx(\LamR )}^{\Lam_0}$.

As a result, the family of limit-point PMs $\{\mu^{\Lam_0}:\;\Lam_0\subset\bbR^d\}$
has the compatibility property and therefore satisfies the assumptions
of the Kolmogorov theorem. This implies that there exists a unique
PM $\mu$ on $(\cW^* (\bbR^d),\fW (\bbR^d))$ such that the restriction of $\mu$ on
the sigma-algebra $\fW (\Lam_0)$ coincides with $\mu^{\Lam_0}$.

The fact that $\mu$ is an FK-DLR PM follows from the above construction.
Hence, each limit-point state $\vphi$
falls in class $\fF_+ (z,\beta )$.
This completes the proof of Theorem 1.2.

\medskip

{\bf Remark.} In the course of the proof of compactness of measures
$\mu^{\Lam_0}_{\Lam |\bx(\Lam R)}$ we did not use the condition (1.1.20).

\section{The shift-invariance of an FK-DLR PM in a plane }

In this section we establish the following theorem (cf. Theorem  1.2.${\bII}$).
\medskip

{\bf Theorem 3.1.} {\sl In dimension two ($d=2$), any FK-DLR PM $\mu\in\fK$ is
translation invariant:
$\forall$  $s=({\ts}^1,{\ts}^2)\in\bbR^2$, square $\Lam_0=[-L_0,L_0]^{\times 2}$  and
event $\cD\in\cW^* (\bbR^2)$ localised in $\Lam_0$ (i.e., belonging to a sigma-algebra
$\fW (\Lam_0)$; cf Definition {\rm{2.4.{\bfI}}}), we have that
$$\mu (\tS (s)\cD)=\mu (\cD).$$
Here $\tS (s)\cD$ stands for the shifted event localised in the shifted square\\
$\tS (s)\Lam_0=[-L_0+\ts^1,\ts^1+L_0]\times [-L_0+\ts^2,\ts^2+L_0]$.$\qquad\Box$}

\medskip

Our Theorem 1.2 is a direct corollary of Theorem 3.1. As in \cite{SKS}, the principal step in the proof of
Theorem 3.1 is

\medskip

{\bf Theorem 3.2.} {\sl Let $\mu$ be an FK-DLR PM, $\Lam_0$ be a square $[-L_0,L_0]^{\times 2}$
and an event $\cD\subset\cW^* (\bbR^2)$ be given, localized in $\Lam_0$:
$\cD\in \fW(\Lam_0)$. Then
$$\mu (\tS(s)\cD)+\mu (\tS(-s)\cD)-2\mu (\cD)\geq 0.\qquad\Box\eqno (3.1)$$}
\medskip

Cf. Theorem 2.1.${\bII}$. The proof of Theorem 3.2 is basically a repetition of that of
Theorem 2.1.${\bII}$ (its main ideas go back to
\cite{R1}--\cite{R3}, particularly \cite{R2}).
Consequently, we will omit various technical details referring the reader to the above
publications. Let $L>L_0$ be given, and set $\Lam =[-L,L]\times [-L,L]$.  The main ingredient
of the proof is a family of maps $\tT^{\,\pm}_L=\tT^{\,\pm}_{L,L_0}(s):\cW^*(\bbR^2)$
$\to \cW^* (\bbR^2)$, $s=(\ts^1,\ts^2)$, featuring properties (i)--(vi) listed in
Sect. 2.${\bII}$. The formal definition of maps $\tT^{\pm}_L$ follows Sect 3.${\bII}$ and is
given in terms of $\ttt$-sections of LCs $(\bOm^*_{\Lam }\vee\bOm^*_{\Lamc})$.
As in \cite{SKS}, Theorem 3.2 can be deduced from Theorem 3.3:

\medskip

{\bf Theorem 3.3.} {\sl For any $\delta >0$ there exists $L^*_0=L^*_0(\delta )>0$ such that
for $L\geq L^*_0$ there exists a subset $\cG_L\subset\cW^*(\bbR^2)$ such that $\cG_L\in\fM$
and the following properties are satisfied:
$$\begin{array}{l}\diy{\rm{(A)}}\qquad\qquad\qquad\quad \mu (\cG_L)=\int_{\cW^* (\bbR^2)}\mu (\rd\bOm^*_{\Lamc})
\;{\mathbf 1}\Big(\bOm^*_{\Lamc}\in\cW^* (\Lamc)\Big)\\
\qquad\diy\times
\int_{\cW^* (\Lam )}\rd\bOm^*_\Lam\;{\mathbf 1}\Big(\bOm^*_\Lam\vee\bOm^*_{\Lamc}
\in \cG_L\Big)\\
\qquad\qquad\qquad\qquad\diy\times\;\frac{z^{K(\bOm^*_\Lam)}}{L(\bOm^*_\Lam)}
\;\exp\,\big[-h(\bOm^*_{\Lam }|\bOm^*_{\Lamc})\,\big]\geq 1-\delta .
\end{array}\eqno (3.2)$$

{\rm{(B)}} The probabilities $\mu (\tS(\pm s)(\cD\cap \cG_L))$ are represented in the form
$$\begin{array}{c}\mu (\tS(\pm s)(\cD\cap \cG_L))=\diy\int_{\cW^*_\td(\bbR^2)}
\mu (\rd\bOm^*_{\Lamc})
{\mathbf 1}\Big(\bOm^*_{\Lamc}\in\cW_\td\left(\Lamc \right)\Big)\\
 \diy\times\int_{\cW^*_\td(\Lam )}
\rd\bOm^*_{\Lam}\;{\mathbf 1}\Big(\bOm^*_\Lam\vee\bOm^*_{\Lamc}\in \cG_L\cap\cD\Big)\;
\frac{z^{K(\bOm^*_\Lam)}}{L(\bOm^*_\Lam)}\\
\qquad\qquad\qquad\times J^\pm_L(\bOm^*_\Lam\vee\bOm^*_{\Lamc})
\exp\,\big[-h(\tT^\pm_L(s)\bOm^*_{\Lam }|\bOm^*_{\Lamc})\,\big]
\end{array}\eqno (3.3)$$
where functions $J^\pm_L=J^\pm_{L,s}$ give the Jacobians of maps $\tT^{\,\pm}_L(s)$.

{\rm{(C)}} Furthermore, the following properties hold true:
$\forall$ $\bOm^*_\Lam\in\cW^* (\Lam )$,
$\bOm^*_{\Lamc}\in\cW^* (\Lamc )$ with $\bOm^*_\Lam\vee
\bOm^*_{\Lamc}\in \cG_L$,
$$\Big[J^+_L(\bOm^*_\Lam\vee\bOm^*_{\Lamc})
J^-_L(\bOm^*_\Lam\vee\bOm^*_{\Lamc})\Big]^{1/2}\geq 1-\delta ;\eqno (3.4{\rm a})$$
and
$$h(\tT^+_L(s)\bOm^*_{\Lam }|\bOm^*_{\Lamc})
+h(\tT^-_L(s)\bOm^*_{\Lam }|\bOm^*_{\Lamc})-
2h(\bOm^*_{\Lam }|\bOm^*_{\Lamc})\leq \delta. \eqno (3.4{\rm b})$$}
$\Box$

{\bf Remark.} As in \cite{SKS}, the dimension $2$ is crucial for properties (3.4a,b).
Theorem 3.3 is the only place where condition ${\ov V}^{(2)}<+\infty$ is used. Cf. (1.1.4).
\medskip

Theorem 3.2 is deduced from Theorem 3.3 in a standard fashion (see Eqns
(2.10.${\bII}$)--(2.12.${\bII}$)).
\medskip

The proof of Theorem 3.2 goes in parallel with that of Theorem 2.2.${\bII}$; a particular
role is played by a specific form of the Jacobians $J^\pm_L(\Om^*_\Lam\vee\bOm^*_{\Lamc})$;
cf. Eqn (3.23.${\bII}$). Here we mark the places where the proof of Theorem 2.2.${\bII}$
(see Sects 3.${\bII}$--5.${\bII}$) has
to be modified, because of the  assumption of non-negativity for the potentials $V_{j,j^\prime}$
and the condition
that fugacities $z_j\in (0,1)$, $1\leq j\leq q$. (a) Every time we use
the Ruelle bound (cf. Eqns (3.27.${\bII}$), (4.12.${\bII}$),
(4.21.${\bII}$)),  we should employ $z_j$ instead of $\;\orho\;$ (defined in Eqn (1.1.4.${\bII}$). (b)
The quantity $\td$ appearing in Eqns (3.13.${\bII}$), (4.4.${\bII}$), (4.5.${\bII}$),
(4.8.${\bII}$), (4.9.1.${\bII}$), (4.9.2.${\bII}$), (4.10.${\bII}$)
(4.13.${\bII}$), (4.14.${\bII}$), (4.17.${\bII}$), (4.19.1.${\bII}$), (4.19.2.${\bII}$). (4.20.1.${\bII}$)\\
(4.20.2.${\bII}$), (4.21.${\bII}$), (5.8.${\bII}$) and (5.9.${\bII}$) should be set to be $0$.

{\bf Acknowledgments.}
This work has been done under Grant 2011/20133-0 provided by
the FAPESP, Grant 2011.5.764.35.0 provided by The Reitoria of the
Universidade de S\~{a}o Paulo (USP) and Grant 11/51845-5 provided by FAPESP. The authors
thank NUMEC and IME-USP,
Brazil, for the warm hospitality.

\end{document}